\documentclass[twocolumn]{revtex4-1}

\usepackage{latexsym}
\usepackage{amssymb}
\usepackage{amsmath}
\usepackage{amsthm}
\usepackage{amsfonts}
\usepackage{ifthen}
\usepackage{revsymb}
\usepackage{yfonts}
\usepackage{graphicx}
\usepackage{algpseudocode}
\usepackage{bbm}
\usepackage{multirow}
\usepackage{gensymb}
\usepackage{epstopdf}
\usepackage{color}
\usepackage{afterpage}
\usepackage{verbatim}
\usepackage{soul}
\usepackage{lineno, blindtext}
\usepackage{footnote}
\usepackage{afterpage}
\usepackage{algpseudocode}
\usepackage[normalem]{ulem}
\usepackage{algorithm}
\usepackage{algpseudocode}

\usepackage[colorlinks = true]{hyperref}
\usepackage{longtable}

\usepackage{xcolor}
\definecolor{darkred}  {rgb}{0.5,0,0}
\definecolor{darkblue} {rgb}{0,0,0.5}
\definecolor{darkgreen}{rgb}{0,0.5,0}

\hypersetup{
  urlcolor   = blue,         
  linkcolor  = darkblue,     
  citecolor  = darkgreen,    
  filecolor   = darkred       
}

\newcommand{\be}{\begin{equation}}
\newcommand{\ee}{\end{equation}}
\newcommand{\bq}{\begin{eqnarray}}
\newcommand{\eq}{\end{eqnarray}}
\newcommand{\bea}{\begin{eqnarray}}
\newcommand{\eea}{\end{eqnarray}}
\newcommand{\ba}{\begin{align}}
\newcommand{\ea}{\end{align}}

\newcommand{\1}{\mathbbm{1}}

\newcommand{\ket}[1]{ | \, #1 \rangle}
\newcommand{\bra}[1]{ \langle #1 \,  |}

\newcommand{\braket}[2]{\left\langle\, #1\,|\,#2\,\right\rangle}

\newcommand{\proj}[1]{\ket{#1}\bra{#1}}

\newcommand{\avr}[1]{\left \langle#1 \right \rangle}

\newcommand{\tr}[1]{{\rm tr}\left[{#1}\right]}

\newcommand{\raw}{\rightarrow}

\newcommand{\bR}{\mathbbm{R}}

\newcommand{\bC}{\mathbbm{C}}
\newcommand{\bZ}{\mathbbm{Z}}

\newcommand{\cH}{\mathcal{H}}

\newcommand{\cU}{\mathcal{U}}

\newcommand{\cS}{\mathcal{S}}

\newcommand{\cM}{\mathcal M}

\newcommand{\cO}{\mathcal O}

\newcommand{\cP}{\mathcal P}

\newcommand{\Sp}{\,\,\,\,\,\,}
\newcommand{\no}{\nonumber\\}

\newcommand{\beginsupplement}{%
        \setcounter{table}{0}
        \renewcommand{\thetable}{S\arabic{table}}%
        \setcounter{figure}{0}
        \renewcommand{\thefigure}{S\arabic{figure}}%
     }

\definecolor{myexpcolor}{RGB}{139,0,139}
\definecolor{myinnercolor}{RGB}{102,102,0}
\definecolor{mygray}{gray}{0.6}

\begin{document}

\title{Supervised learning with quantum enhanced feature spaces}

\author{Vojtech Havlicek$^1$}
\thanks{On leave from Quantum Group, Department of Computer Science, University of Oxford, Wolfson Building, Parks Road, Oxford OX1 3QD, UK}
\author{Antonio D. C\'orcoles$^1$}
\author{Kristan Temme$^1$}
\author{\mbox{Aram W. Harrow}$^2$}
\author{Abhinav Kandala$^1$}
\author{Jerry M. Chow$^1$}
\author{Jay M. Gambetta$^1$}

\affiliation{$^1$IBM T.J.  Watson  Research  Center,  Yorktown  Heights,  NY 10598,  USA}
\affiliation{$^2$Center for Theoretical Physics, Massachusetts Institute of Technology, USA}

\date{\today}

\maketitle
\noindent
{\bf Machine learning and quantum computing are two technologies each   with the potential for altering how computation is performed to  address previously untenable problems. Kernel methods for machine learning are ubiquitous for pattern recognition, with support vector machines (SVMs) being the most well-known method for classification problems. However, there are limitations to the successful solution to such problems when the feature space becomes large, and the kernel functions become computationally expensive to estimate. A core element to computational speed-ups afforded by quantum algorithms is the exploitation of an exponentially large quantum state space through controllable entanglement and interference. 

Here, we propose and experimentally implement two novel methods on a superconducting processor. Both methods represent the feature space of a classification problem by a quantum state, taking advantage of the large dimensionality of quantum Hilbert space to obtain an enhanced solution. One method, the quantum variational classifier builds on \cite{mitarai2018quantum,farhi2018classification} and operates through using a variational quantum circuit to classify a training set in direct analogy to conventional SVMs. In the second, a quantum kernel estimator, we estimate the kernel function and optimize the classifier directly. The two methods present a new class of tools for exploring the applications of noisy intermediate scale quantum computers \cite{preskill2018quantum} to machine learning. \\}

The intersection between machine learning and quantum computing has been dubbed quantum machine learning, and has attracted considerable attention in recent years \cite{Arunachalam:2017:GCS,ciliberto2018quantum,dunjko2018machine}. This has led to a number of recently proposed quantum algorithms  \cite{biamonte2017quantum,romero2017quantum,wan2016quantum,mitarai2018quantum,farhi2018classification}. Here, we present a  quantum algorithm that has the potential to run on near-term quantum devices. A natural class of algorithms for such noisy devices are short-depth circuits, which are amenable to error-mitigation techniques that reduce the effect of decoherence \cite{temme2017error,li2017efficient}. There are convincing arguments that indicate that even very simple circuits are hard to simulate by classical computational means \cite{terhal2002adaptive,Bremner2017achievingsupremacy}. The algorithm we propose takes on the original problem of supervised learning: the construction of a classifier. For this problem, we are given data from a training set $T$ and a test set $S$ of a subset $\Omega \subset \bR^d$. Both are assumed to be labeled by a map $m: T \cup S \rightarrow \{+1,-1\}$ unknown to the algorithm. The training algorithm only receives the labels of the training data $T$. The goal is to infer an approximate map on the test set $\tilde{m}: S \rightarrow \{+1,-1\}$ such that it agrees with high probability with the true map $m(\vec{s}) = \tilde{m}(\vec{s})$ on the test data $\vec{s} \in S$. For such a learning task to be meaningful it is assumed that there is a correlation between the labels given for training and the true map.  A classical approach to constructing an approximate labeling function uses so-called support vector machines (SVMs) \cite{vapnik2013nature}. The data gets mapped non-linearly to a high dimensional space, {\it the feature space}, where a hyperplane is constructed to separate the labeled samples. A quantum version of this approach has already been proposed in \cite{rebentrost2014quantum}, where an exponential improvement can be achieved if data is provided in a coherent superposition. However, when data is provided in the conventional way, i.e. from a classical computer, then the methods of \cite{rebentrost2014quantum} cannot be applied. 

\begin{figure*}
	\begin{center}
		\includegraphics[width=0.9\textwidth]{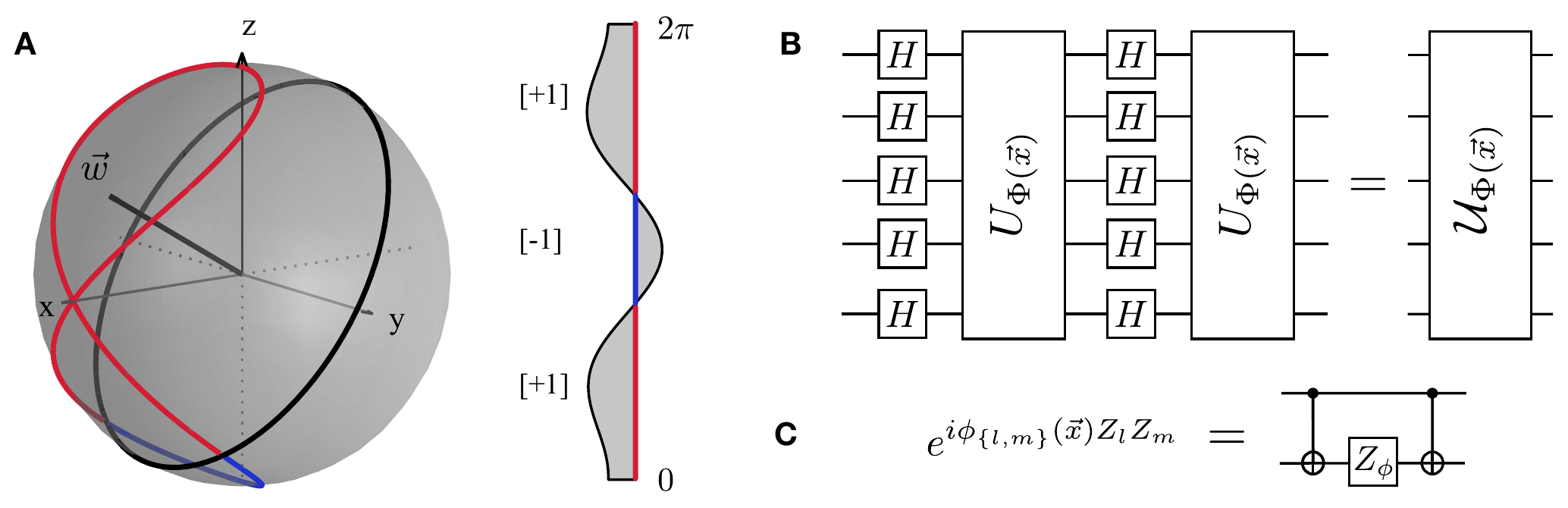}
		\caption{\label{Figure1} \textbf{Quantum Kernel Functions}: (a) Feature map representation for a single qubit. A classical dataset in the interval $\Omega = (0, 2\pi]$ with binary labels (a, right) can be mapped onto the Bloch sphere (red / blue - lines) by using the non-linear feature map described in (b). For a single qubit $U_{{\Phi}(x)} = Z_x$ is a  phase-gate of angle $x \in \Omega$. The mapped data can be separated by the hyperplane given by normal $\vec{w}$. States with a positive expectation value of $\vec{w}$ receive a $[+1]$ (red) label, while negative values are labeld $[-1]$(blue). (b) For the general circuit $U_{{\Phi}(\vec{x})}$ is formed by products of single- and two-qubit unitaries that are diagonal in the computational basis. In our experiments, both the training and testing data is artificially generated to be perfectly classifiable with the aforementioned feature map. The circuit family depends non-linearly on the data through the coefficients $\phi_S(\vec{x})$ with $|S| \leq 2$. (c) Experimental implementation of the parameterized diagonal two-qubit operations using CNOTs and $Z-$gates.}
	\end{center}
\end{figure*}

Here, we propose two SVM type classifiers that process data provided purely classically and use the quantum state space as the feature space to still obtain a quantum advantage. This is done by mapping the data non-linearly to a quantum state  $\Phi: \vec{x} \in \Omega \raw \proj{\Phi(\vec{x})}$, c.f. Fig ~\ref{Figure1}(a). We implement both classifiers on a superconducting quantum processor. In the first approach we use a variational  circuit as given in \cite{kandala2017hardware,farhi2017quantum,farhi2018classification,mitarai2018quantum} that generates a separating hyperplane in the quantum feature space. In the second approach we use the quantum computer to estimate the kernel function of the quantum feature space directly and implement a conventional SVM.  A necessary condition to obtain a quantum advantage, in either of the two approaches, is that the kernel cannot be estimated classically.  This is true, even when complex variational quantum circuits are used as classifiers. In the experiment, we want to disentangle the question of whether the classifier can be implemented in hardware, from the problem of choosing a suitable feature map for a practical data set. The data that is classified here is chosen so that it can be classified with $100\%$ success to verify the method. We demonstrate that this success ratio is subsequently achieved in experiment.

\begin{figure}
\begin{center}
\includegraphics[width=3.4in]{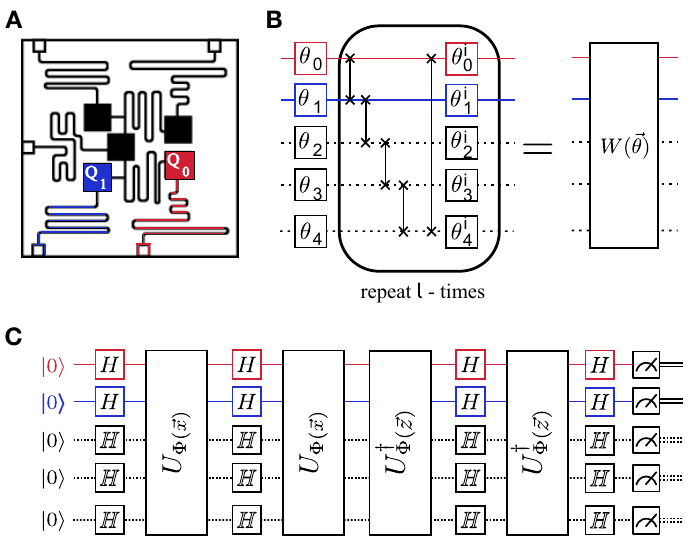}
\caption{\label{Figure2} \textbf{Experimental implementations} (a) Schematic of the 5-qubit quantum processor. The experiment was performed on qubits $Q_0$ and $Q_1$, highlighted in the image. (b) Variational circuit used for our optimization method. Here we choose a rather common Ansatz for the variational unitary $W(\vec{\theta}) = U_{\textrm{loc}}^{(l)}(\theta_l) \;U_{ent} \ldots U_{\textrm{loc}}^{(2)}(\theta_2) \; U_{\textrm{ent}}\; U_{\textrm{loc}}^{(1)}(\theta_1)$ \cite{kandala2017hardware,farhi2017quantum}.  We alternate layers of entangling gates $U_{ent}= \prod_{(i,j) \in E} \textsf{CZ}(i,j)$ with full layers single qubit rotations $U_{loc}^{(t)}(\theta_t) = \otimes_{i=1}^n U(\theta_{i,t})$ with  $U(\theta_{i,t}) \in \text{SU}(2)$. For the entangling step we use controlled phase gates $\textsf{CZ}(i,j)$ along the edges $(i,j) \in E$ present in the connectivity of the superconducting chip. (c) Circuit to directly estimate the fidelity between a pair of feature vectors for data $\vec{x}$ and $\vec{y}$ as used for our second method.}
\end{center}
\end{figure}

Our experimental device consists of five coupled superconducting transmons, only two of which are used in this work, as shown in Fig.~\ref{Figure2}(a).  Two co-planar waveguide (CPW) resonators, acting as quantum buses, provide the device connectivity. Each qubit has one additional CPW resonator for control and readout. Entanglement in our system is achieved via CNOT gates, which use cross-resonance \cite{Rigetti2010} as well as single qubit gates as primitives. The quantum processor is thermally anchored to the mixing chamber plate of a dilution refrigerator.

{\it Quantum feature map:} Before discussing the two methods of classification, we discuss the feature map. Training and classification with conventional support vector machines is efficient when inner products between feature vectors can be evaluated efficiently \cite{vapnik2013nature,burges1998tutorial,boser1992training}. We will see that classifiers based on quantum circuits, such as the one presented in Fig~\ref{Figure2}(c) cannot provide a quantum advantage over a conventional support vector machine if the feature vector kernel $K(\vec{x},\vec{z}) = |\braket{\Phi(\vec{x})}{\Phi(\vec{z})}|^2$ is too simple. For example, a classifier that uses a feature map that only generates product states can immediately be implement classically. To obtain an advantage over classical approaches we need to implement a map based on circuits that are hard to simulate classically.  Since quantum computers are not expected to be classically simulable, there exists a long list of (universal) circuit families one can choose from. Here, we propose to use a circuit that works well in our experiments and is not too deep. We define a feature map on $n$-qubits generated by the unitary $\cU_{\Phi}(\vec{x}) = U_{\Phi(\vec{x})} H^{\otimes n}  U_{\Phi(\vec{x})} H^{\otimes n}$, where $H$ denotes the conventional Hadamard gate and
\begin{equation*}\label{Fearure_map}
U_{\Phi(\vec{x})} = \exp\left(i \sum_{S \subseteq [n]} \phi_S(\vec{x}) \prod_{i \in S} Z_i\right),
\end{equation*}
is a diagonal gate in the Pauli $Z$ - basis, c.f. Fig~\ref{Figure1} (b). This circuit will act on $\ket{0}^n$ as initial state. We use the coefficients $\phi_S(\vec{x}) \in \bR$, to encode the data $\vec{x} \in \Omega$.
In general any diagonal unitary $U_{\Phi(\vec{x})}$ can be used if it can be implemented efficiently. This is for instance the case when only weight $|S| \leq 2$ interactions are considered. The exact evaluation of the inner-product between two states generated from a similar circuit with only a single diagonal layer $U_{\Phi(\vec{x})}$ is $\#P$ - hard  \cite{goldberg2017complexity}. Nonetheless, in the experimentally relevant context of additive error approximation, simulation of a single layer preparation circuit can be achieved efficiently classically by uniform sampling \cite{demarie2018classical}. We conjecture that the evaluation of inner products generated from circuits with two basis changes and diagonal gates up to additive error to be hard, c.f. supplementary material for a discussion.\\

\begin{figure*}
	\begin{center}
		\includegraphics[width=0.8\textwidth]{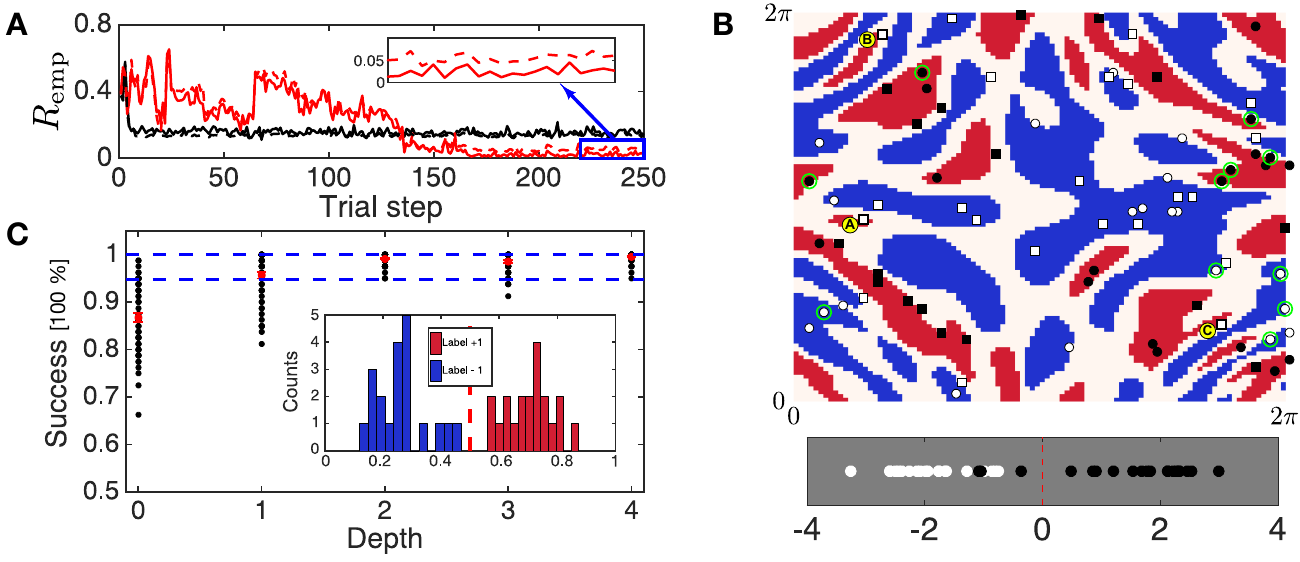}
		\caption{\label{Figure3} \textbf{Convergence of the method and classification results}: (a) Convergence of the cost function $R_{\textrm{emp}}(\vec{\theta})$ after $250$ iterations of Spall's SPSA algorithm. Red (black) curves correspond to $l=4$ ($l=0$). The value of the cost function, obtained from estimates of $\hat{p}_k$ after zero-noise extrapolation (solid lines), is compared with values obtained from experimental measurements of $\hat{p}_k$ with standard single and two-qubit gate times (dashed).  We train three sets of data per depth and perform $20$ classifications per trained set. The results of these classifications are shown in (c) as black dots (amounting to 60 per depth), with mean values at each depth represented by red dots. The error bar is the standard error of the mean. The inset shows histograms as a function of the probability of measuring label $+1$ for a test set of $20$ points per label obtained with a $l=4$ classifier circuit, depicting classification with $100\%$ success.. The dashed blue lines show the results of our direct kernel estimation method for comparison, with Sets I and II yielding $100\%$ success and Set III yielding $94.75\%$ success. (b) Example data set used for both methods in this work. The data labels (red for $+1$ label and blue for $-1$ label) are generated with a separation gap of magnitude 0.3 between them (white areas). The 20 points per label training set is shown as white and black circles. For our quantum kernel estimation method we show the obtained support vectors (green circles) and a classified test set (white and black squares). Three of the test sets points are misclasified, labeled as A, B, and C. For each of the test data points $\vec{x}_j$ we plot at the bottom of (b) the amount $\sum_i y_i \alpha^*_i K(\vec{x}_i, \vec{x}_j)$ over all support vectors $\vec{x}_i$, where $y_i \in \{+1, -1\}$ are the labels of the support vectors. Points A, B, and C, all belonging to label $+1$, give $\sum_i y_i \alpha^*_i K(\vec{x}_i, \vec{x}_j) =$  -1.033, -0.367 and -1.082, respectively.}
	\end{center}
\end{figure*}

{\it The data:} To test our two methods, we generate artificial data that can be fully separated by our feature map. We use the map for $n = d = 2$ - qubits in Fig.~\ref{Figure1}(b) with  $\phi_{\{i\}}(\vec{x}) = x_i$ and $\phi_{\{1,2\}}(\vec{x}) = (\pi-x_1)(\pi-x_2)$. We generate the labels for data vectors $\vec{x} \in T \cup S \subset (0,2\pi]^2$, by first choosing ${\bf f} =  Z_1 Z_2$ as the parity function and a random unitary $V \in SU(4)$. We assign $m(\vec{x})  = +1$, when $\bra{\Phi(\vec{x})} V^\dag {\bf f} V \ket{\Phi(\vec{x})} \geq \Delta$ and $m(\vec{x})  = -1$ when $\bra{\Phi(\vec{x})} V^\dag {\bf f} V \ket{\Phi(\vec{x})} \leq -\Delta$, c.f. Fig~\ref{Figure3}(b). The data has been separated by a gap of $\Delta = 0.3$. Both the training sets and the classification sets consist of $20$ data points per label. We show one of such classification sets as circle symbols in Fig.~\ref{Figure3}(b). \\

{\it Quantum variational classification:} The first classification protocol follows four steps. First, the data $\vec{x} \in \Omega$ is mapped to a quantum state by applying the feature map circuit $\cU_{\Phi(\vec{x})}$ in Fig.~\ref{Figure1}(b) to a reference state $\ket{0}^n$. Second, a short depth quantum circuit $W(\vec{\theta})$, described in Fig ~ \ref{Figure2}(b) is applied to the feature state. A circuit with $l$ - layers is parametrized by $\vec{\theta} \in \bR^{2 n (l+1)}$ that will be optimized during training. Third, for a two label classification ${y \in \{+1,-1\}}$ problem, a binary measurement $\{ M_y\}$ is applied to the state $W(\vec{\theta}) \cU_{\Phi(\vec{x})}\ket{0}^n$. This measurement is implemented by measurements in the $Z$ - basis and feeding the output bit-string $z \in \{0,1\}^n$ to a chosen boolean function $f : \{0,1\}^n \raw \{+1,-1\}$. The measurement operator is given by $M_{y} = 2^{-1}(\1 + y {\bf f})$, where we have defined  ${\bf f} = \sum_{z \in \{0,1\}^n} f(z) \proj{z}$. The probability of obtaining outcome $y$ is $p_y(\vec{x}) = \bra{\Phi(\vec{x})} W^\dag(\vec{\theta}) M_y W(\vec{\theta}) \ket{\Phi(\vec{x})}$. Fourth, for the decision rule we perform $R$ repeated measurement shots to obtain the empirical distribution $\hat{p}_y(\vec{x})$. We assign the label $\tilde{m}(\vec{x}) = y$, whenever $\hat{p}_y(\vec{x}) > \hat{p}_{-y}(\vec{x}) - yb$, where we have introduced an additional bias parameter $b \in [-1,1]$ that can be optimized during training. 

The feature map circuit $\cU_{\Phi(\vec{x})}$ as well as the boolean function $f$ are fixed choices. During the training of the classifier we optimize the parameters $(\vec{\theta},b)$. For the optimization, we need to define a cost-function. We define the empirical risk $R_{\textrm{emp}}(\vec{\theta})$ given by the error probability $\text{Pr} \left( \tilde{m}(\vec{x}) \neq m(\vec{x})\right)$ of assigning the incorrect label averaged over the samples in the training set $T$,
\begin{align*}
R_{\textrm{emp}}(\vec{\theta}) &=  \frac{1}{|T|}  \sum_{\vec{x} \in T}\text{Pr} \left( \tilde{m}(\vec{x}) \neq m(\vec{x})\right)\,.
\end{align*}
For the binary problem, the error probability of assigning the wrong label is given by the binomial cumulative density function (CDF) of the empirical distribution $\hat{p}_y(\vec{x})$, c.f. supplementary material for a derivation. The binomial CDF can be approximated for a large number of samples (shots) $R \gg 1$ by a sigmoid function $\mbox{sig}(x) = (1 + e^{-x})^{-1}$. The probability that the label $m(\vec{x}) = y$ is assigned incorrectly is approximated by
\begin{align*}
\text{Pr} \left( \tilde{m}(\vec{x}) \neq m(\vec{x})\right) \approx \mbox{sig} \left(\frac{\sqrt{R} \left(\frac{1}{2} - \left(\hat{p}_y(\vec{x}) - \frac{y b}{2} \right) \right)}{\sqrt{2(1 - \hat{p}_y(\vec{x}))\hat{p}_y(\vec{x})}}\right).
\end{align*}
The experiment itself is split in to two phases; First, we train the classifier and optimize $(\vec{\theta},b)$. We have found that Spall's SPSA \cite{OneSPSA,AdaptiveSPSA} stochastic gradient decent algorithm performs well in the noisy experimental setting. We can use the circuit as a classifier after the parameters have converged to $(\vec{\theta}^*,b^*)$. Second, in the classification phase, the classifier assigns labels to unlabeled data $\vec{s} \in S$ according to the decision rule $\tilde{m}(\vec{s})$. \\

We implement the quantum variational classifier $W(\vec{\theta})$ over 5 different depths ($l=0$ through $l=4$), c.f. Fig~\ref{Figure2}(b), in our superconducting quantum processor. We expect a higher classification success for increased depth. The binary measurement is obtained from the parity function ${\bf f} = Z_1Z_2$. For each depth we train three different data sets, using training sets consisting of $20$ data points per label. One of these data sets is shown in Fig. \ref{Figure3} (b), along with the training set used for this particular data set.
Fig.~\ref{Figure3}(a) shows the optimization of the empirical risk $R_{\textrm{emp}}(\vec{\theta})$ for two different training sets and depths. In all experiments throughout this work we implemented an error mitigation technique which relies on zero-noise extrapolation to first order \cite{temme2017error,Kandala18}. To obtain a zero-noise estimate, a copy of the circuit was run on a time scale slowed down by a factor of $1.5$, c.f. supplemental material. This technique is implemented at each trial step, and it is the mitigated cost function that is fed to the classical optimizer. We observe that the empirical risk in Fig.~\ref{Figure3}(a) converges to a lower value for depth $l=4$ than for $l=0$, albeit with more optimization steps. Whereas error mitigation does not appreciably improve the results for depth 0 - the noise in our system is not the limiting factor in that case-, it does help substantially for larger depths. Although $\text{Pr} \left( \tilde{m}(\vec{x}) \neq m(\vec{x})\right)$  explicitly includes the number of experimental shots taken, we fixed $R = 200$ to avoid gradient problems, even though we took $2000$ shots in the actual experiment.

After each training is completed, we use the trained set of parameters $(\vec{\theta}^*, b^* = 0)$ to classify 20 different test sets -randomly drawn each time- per data set. We run these classification experiments at 10,000 shots, versus the 2,000 used for training. The classification of each data point is error-mitigated and repeated twice, averaging the success ratios obtained in each of the two classifications. Fig.~\ref{Figure3} (c) shows the classification results for our quantum variational approach. We clearly see an increase in classification success with increasing circuit depth, c.f. Fig.~\ref{Figure3}(c), reaching values very close to $100\%$ for depths larger than 1. This classification success remarkably remains up to depth 4, despite the decoherence associated with 8 CNOTs in the training and classification circuits, for $l=4$.\\

{\it A path to quantum advantage:} Such variational circuit classifiers are directly related to conventional SVMs \cite{vapnik2013nature,burges1998tutorial}. To see why a quantum advantage can only be obtained for feature maps with a classically hard to estimate kernel, we point out the following:  The decision rule $p_y(\vec{x}) > p_{-y}(\vec{x}) - yb$ can be restated as $\tilde{m}(\vec{x}) = \mbox{sign}( \bra{\Phi(\vec{x})} W^\dag(\vec{\theta}) {\bf f} W(\vec{\theta}) \ket{\Phi(\vec{x})} + b)$. The variational circuit $W$ followed by a binary measurement can be understood as a separating hyperplane in quantum state space. Choose an orthogonal, hermitian, matrix basis $\{P_\alpha\} \subset \bC^{2^n \times 2^n}$, where $\alpha = 1,\ldots,4^n$ with $\tr{P^\dag_\alpha P_\beta} = 2^n \delta_{\alpha,\beta}$ such as the Pauli-group on $n$-qubits. Expand both the quantum state  $\proj{\Phi(\vec{x})}$ and the measurement $W^\dag(\vec{\theta}) {\bf f} W(\vec{\theta})$ in this matrix basis. Both the expectation value of the binary measurement and the decision rule can be expressed in terms of  $w_\alpha(\vec{\theta})  = \tr{W^\dag(\vec{\theta}) {\bf f} W(\vec{\theta}) P_\alpha}$ and $ \Phi_\alpha(\vec{x}) = \bra{\Phi(\vec{x})} P_\alpha \ket{\Phi(\vec{x})}$. For any variational unitary the classification rule can be restated in the familiar SVM form $\tilde{m}(x) = \mbox{sign}\left(2^{-n} \sum_{\alpha} w_\alpha(\vec{\theta}) \Phi_\alpha(\vec{x}) + b \right)$. The classifier can only be improved when the constraint is lifted that the $w_\alpha$ come from a variational circuit. The optimal $w_\alpha$ can alternatively be found by employing kernel methods and considering the standard Wolfe - dual of the SVM \cite{vapnik2013nature}. Moreover, this decomposition indicates that one should think of the feature space as the quantum state space with feature vectors $\proj{\Phi(\vec{x})}$ and inner products  $K(\vec{x},\vec{z}) =  |\braket{\Phi(\vec{x})}{\Phi(\vec{z})}|^2$. Indeed, the direct use of the Hilbert space $\cH = (\bC^2)^{\otimes n}$ as a feature space would lead to a conceptual problem, since a vector $\ket{\Phi(\vec{x})} \in \cH$ is only physically defined up to a global phase.\\

{\it Quantum kernel estimation:} The second classification protocol uses this connection to implement the SVM directly. Rather than using a variational quantum circuit to generate the separating hyperplane, we use a classical SVM for classification. The quantum computer is used twice in this protocol. First, the kernel $K(\vec{x}_i,\vec{x}_j)$  is estimated on a quantum computer for all pairs of training data $\vec{x}_i,\vec{x}_j \in T$, c.f. Fig.~\ref{Figure2}(c). Here it will be convenient to write $T =\{\vec x_1,\ldots,\vec x_t\}$ with $t=|T|$; also let $y_i = m(\vec{x}_i)$ be the corresponding label. The optimization problem for the optimal SVM can be formulated in terms of a dual quadratic program that only uses access to the kernel. We maximize
\begin{equation*}
  L_D(\alpha) = \sum_{i=1}^t\alpha_i - \frac{1}{2}  \sum_{i,j=1}^t y_iy_j \alpha_i \alpha_j K(\vec{x}_i,\vec{x}_j),
\end{equation*}
subject to $ \sum_{i=1}^{t} \alpha_i y_i = 0$ and $\alpha_i \geq 0$ for each $i$. This problem is concave whenever  $K(\vec{x}_i,\vec{x}_j)$ is a positive definite matrix. The solution to this problem will be given by a nonnegative vector $\vec\alpha= (\alpha_1,\ldots,\alpha_t)$. The quantum computer is used a second time to estimate the kernel for a new datum $\vec{s} \in S$ with all the support vectors. The optimal solution $\vec\alpha^*$ is used to construct the classifier 
\begin{equation*}
	\tilde{m}(\vec{s}) = \mbox{sign}\left(\sum_{i=1}^t y_i \alpha^*_i K(\vec{x}_i,\vec{s}) + b\right).
\end{equation*}
Due to complementary slackness, we expect that many of the $\alpha_i$ will be zero.  This can make the evaluation of $\tilde{m}(\vec{s})$ cheaper, since $K(\vec{x}_i,\vec{s})$  only needs to be estimated when $\alpha_i^*>0$. The bias $b$ in $\tilde{m}(\vec{s})$  can calculated from the weights  $\alpha^*_i$ by choosing any $i$ with $\alpha_i^*>0$ and solving $\sum_{j } y_j\alpha^*_j K(\vec{x}_j,\vec{x_i}) + b = y_i$ for $b$.\\ Let us discuss how the quantum computer is used to estimate the kernel. The kernel entries are the fidelities between different feature vectors. Various methods  \cite{buhrman2001quantum,cincio2018learning} exist, such as the swap test, to estimate the fidelity between general quantum states. However, since the states in the feature space are not arbitrary, the overlap can be estimated directly from the transition amplitude $|\braket{\Phi(\vec{x})}{\Phi(\vec{z}}|^2 = |\bra{0^n}\cU^\dag_{\Phi(\vec{x})}\cU_{\Phi(\vec{z})}\ket{0^n} |^2$. First, we apply the circuit Fig.~\ref{Figure2}(c), a composition of two consecutive feature map circuits, to the initial reference state $\ket{0^n}$. Second, we measure the final state in the $Z$-basis $R$ - times and record the number of all zero strings $0^n$. The frequency of this string is the estimate of the transition probability. The kernel entry is obtained to an additive sampling error of $\tilde{\epsilon}$ when $\cO(\tilde{\epsilon}^{-2})$ shots are used. In the training phase a total of $\cO(|T|^2)$ amplitudes have to be estimated. An estimator $\hat{K}$ for the kernel matrix that deviates with high probability  in operator norm from the exact kernel  $K$ by at most $\|K - \hat{K} \| \leq \epsilon$  can be obtained with a total of $R = \cO( \epsilon^{-2} |T|^4)$ shots. The sampling error can compromise the positive semi-definiteness of the kernel.  Although not applied in this work, this can be remedied by employing an adaption of the scheme presented in \cite{smolin2012efficient}. The direct connection to conventional SVMs enables us to use the conventional bounds on the $VC$-dimension that ensure convergence and guide the structural risk minimization. \\

\begin{figure}
	\begin{center}
		\includegraphics[width=3in]{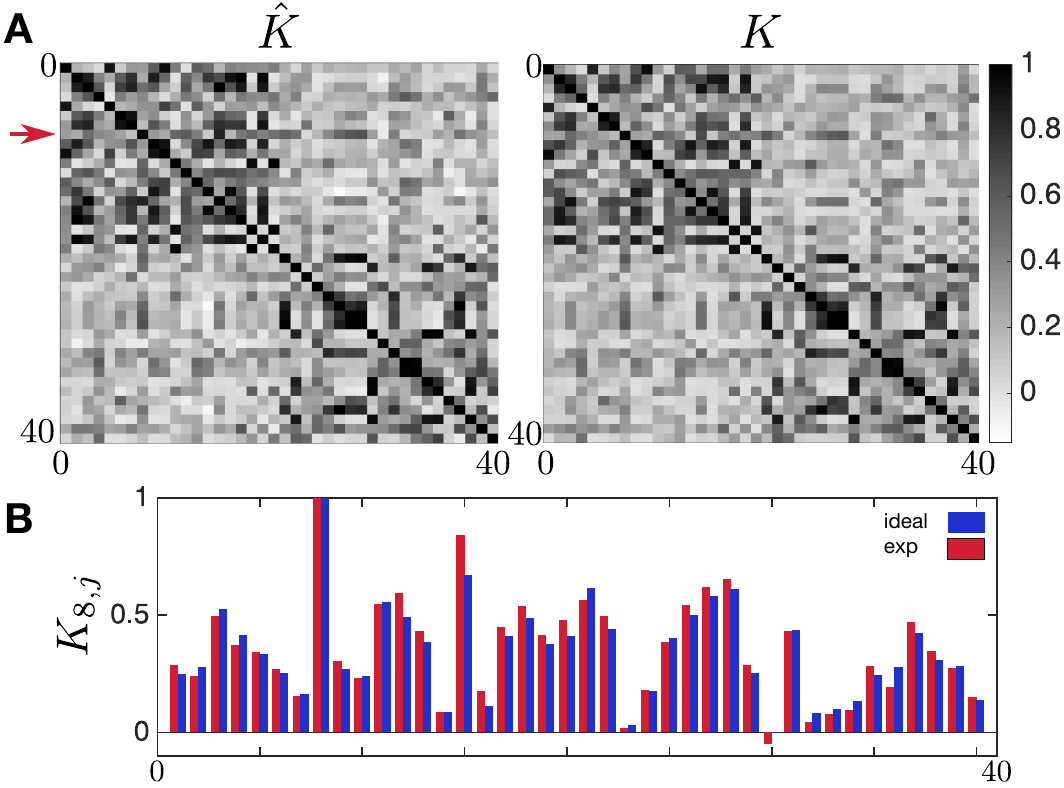}
		\caption{\label{Figure4} \textbf{Kernels for Set III}. (a) Experimental (left) and ideal (right) kernel matrices containing the inner products of all data points used for training Set III (round symbols in Fig. \ref{Figure3} (b)). The maximal deviation from the ideal kernel $|K - \hat{K}|$ occurs at element $K_{8,15}$. A cut through row 8 (indicated by red arrow in (a)) is shown in (b), where the experimental (ideal) results are shown as red (blue) bars. Note that entries that are close zero in the kernel can become negative (c.f. $K_{8,30}$ (b)) when the error-mitigation technique is applied.}
	\end{center}
\end{figure}
For the experimental implementation of estimating the kernel matrix $K$, c.f. circuit Fig.~\ref{Figure2}(c), we again  apply the error-mitigation protocol \cite{temme2017error,Kandala18} to first order. The kernel entries are obtained by running a time-stretched copy of the circuit and reporting the mitigated entry. We use $50,000$ shots per matrix entry. Using this protocol, we obtain support vectors $\alpha_i$ that are very similar to the noise-free case. We run the training stage on three different data sets, which we will label as Set I, Set II and Set III. Set III is shown  in Fig.~\ref{Figure3}(b). Note that the training data used to obtain the kernel and the support vectors is the same data used in training of our variational classifier. The support vectors (green circles in (b)) are then used to classify 10 different test sets randomly drawn from each entire set. Set I and Set II yield $100\%$ success over the classification of all 10 different test sets each, whereas Set III averages a success of $94.75\%$.  For more details see the supplementary information. These classification results are given in Fig.~\ref{Figure3}(c) as dashed blue lines to compare with the results of our variational method. In Fig.~\ref{Figure4}(a) we show the ideal and the experimentally obtained kernel matrices, $K$ and $\hat{K}$, for Set III. The maximum difference across the matrices between $K$ and $\hat{K}$ is found at row (or column) 8. This is shown in Fig.~\ref{Figure4}(b). All support vectors for the three sets and equivalent plots are given in the supplementary material.\\

{\it Conclusions:} We have experimentally demonstrated a  classifier that exploits a quantum feature space. The kernel of this feature space has been conjectured to be hard to estimate classically. In the experiment we find that even in the presence of noise, we are capable of achieving success rates up to 100$\%$. In the future it becomes intriguing to find suitable feature maps for this technique with provable quantum advantages while providing significant improvement on real world data sets. With the ubiquity of kernel methods in machine learning, we are optimistic that our technique will find application beyond binary classification.\\

During composition of this manuscript we became aware of the independent theoretical work by Schuld et al. \cite{schuld2018quantum,schuld2018circuit}.\\

{\bf Supplementary Information} is available in the online version of the paper.\\

{\bf Acknowledgments}
We thank Sergey Bravyi for insightful discussions. A.W.H. acknowledges funding from the MIT-IBM Watson AI Lab under the project  {\it Machine Learning in Hilbert space}. The research was supported by the IBM Research Frontiers Institute. We  acknowledge  support  from  IARPA  under  contract  W911NF-10-1-0324 for device fabrication.\\

{\bf Author contributions}
The work on the classifier theory was led by V.H. and K.T.  The experiment was designed by A.D.C, J.M.G and K.T. and implemented by A.D.C. All authors contributed to the manuscript.\\

{\bf Author information} The authors declare no competing financial interests. Correspondence and requests for materials should be addressed to A.D.C. and K.T.

\pagebreak
\onecolumngrid

\section*{ Supplementary Information: \\
\vspace{0.3cm} Supervised learning with quantum enhanced feature spaces}

\beginsupplement
\subsection*{Classification problems}
Consider a classification task on a set $C =  \lbrace 1, 2 \ldots c \rbrace$ of $c$ classes (labels) in a supervised learning scenario. In such settings, we are given a training set $T$ and a test set $S$, both are assumed to be labeled by a map $m: T \cup S \rightarrow C$ unknown to the programmer. Both sets $S$ and $T$ are provided to the programmer, but the programmer only receives the labels of the training set. So, formally, the programmer has only access to a restriction $m_{|T}$ of the indexing map $m$:
\begin{align*}
m_{|T}: T \rightarrow C, \; \text{s.t.:} \;  m_{|T}(\vec{t}) = m(\vec{t}), \; \forall \vec{t} \in T.
\end{align*} 
It is the programmer's goal to use the knowledge of $m_{|T}$ to infer an indexing map $\tilde{m}: S \rightarrow C$ over the set $S$, such that $m(\vec{s}) = \tilde{m}(\vec{s})$ with high probability for any $\vec{s} \in S$. The accuracy of the approximation to the map is quantified by a classification success rate, proportional to the number of collisions of $m$ and $\tilde{m}$:
\begin{align*}
 \nu_{succ.} &= \frac{|\lbrace \vec{s} \in S | m(\vec{s}) = \tilde{m}(\vec{s}) \rbrace|}{|S|}.
 \end{align*}
For such a learning task to be meaningful it is assumed that there is a correlation in output of the indexing map $m$ over the sets $S$ and $T$. For that reason, we assume that both sets could in principle be constructed by drawing the $S$ and $T$ sample sets from a family of $d$-dimensional distributions $\left \lbrace p_c: \Omega  \subset \mathbb{R}^d \rightarrow \mathbb{R} \right \rbrace_{c \in C}$ and labeling the outputs according to the distribution. It is assumed that the hypothetical classification function $m$ to be learned is constructed this way. The programmer, however, does not have access to these distributions of the labeling function directly. She is only provided with a large, but finite number of  samples and the matching labels.\\

The conventional approach to this problem is to construct a family of classically computable function $\tilde{m}: \langle \vec{\theta}, S \rangle \rightarrow C$, indexed by a set of parameters $\vec{\theta}$. These weights are then inferred from $m_{|T}$ by a optimization procedure on a classical cost function. We consider a scenario where the whole, or parts of the classification protocol $m$, are generated on a quantum computer.

\subsection*{Description of the Algorithm}
We consider two different learning schemes. The first is referred to as ``Quantum variational classification'', the second is referred to ``Quantum kernel estimation''.  Both schemes construct a separating hyperplane in the state space of $n$ qubits. The classical data is mapped to this space with $\mbox{dim} = 4^n$ using a unitary circuit family starting from the reference state $\proj{0}^n$. 
 
\subsubsection*{Quantum variational classification}
\begin{figure}[h]
\begin{center}
\includegraphics[scale=1]{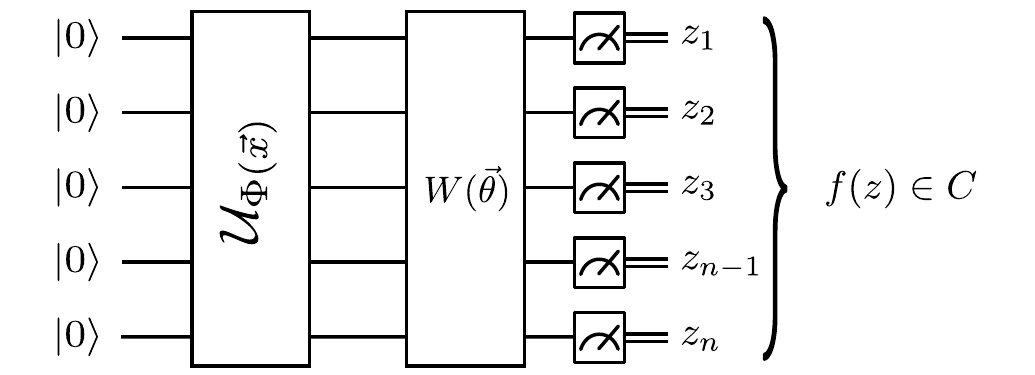}
\caption{\label{fullShortCircuit} Quantum variational classification: The circuit takes a references state, $\ket{0}^n$, applies the unitary ${\cal U}_{\Phi(vec{x})}$ followed by the variational unitary $W(\vec{\theta})$ and applies a measurement in the $Z$-basis. The resulting bit string $z \in \{0,1\}^n$ is then mapped to a label in $C$. This circuit is re - run multiple times and sampled to estimate the expectation value $p_y =  \bra{\Phi(\vec{x})} W^\dag(\vec{\theta}) M_y W(\vec{\theta})\ket{\Phi(\vec{x})}$ for the labels $y \in C$. In the experiment we consider $C = \{+1,-1\}$.}
\end{center}
\end{figure}

For our first classification approach we design a variational algorithm which exploits the large dimensional Hilbert space of our quantum processor to find an optimal cutting hyperplane in a similar vein as Support Vector Machines (SVM) do. The algorithm consists of two main parts: a training stage and a classification stage. For the training stage, a set of labeled data points are provided, on which the algorithm is performed. For the classification stage, we take a different set of data points and run the optimized classifying circuit on them without any label input. Then we compare the label of each data point to the output of the classifier to obtain a success ratio for the data set. For both the training and the classification stages, the quantum circuit that implements the algorithm comprises three main parts: the encoding of the feature map, the variational optimization and the measurement, Fig~\ref{fullShortCircuit}. The training phase consists of these steps.

\begin{algorithm}[H]
\caption{Quantum variational classification: the training phase}
\begin{algorithmic}[1]
\State  {\bf Input} Labeled training samples $T = \{\vec{x} \in \Omega \subset \bR^n\} \times \{y \in C\}$, Optimization routine,
\State  {\bf Parameters} Number of measurement shots $R$, and initial parameter $\vec{\theta}_0$.
\State Calibrate the quantum Hardware to generate short depth trial circuits.
\State Set initial values of the variational parameters $\vec{\theta} = \vec{\theta}_0$ for the short-depth circuit $W(\vec{\theta})$
\While{Optimization (e.g. SPSA) of $R_{\textrm{emp}}(\vec{\theta})$ has not converged}
\For{\texttt{$i=1$ to $|T|$ }}
	\State Set the counter $\mbox{r}_{y} = 0$ for every $y \in C$.
	\For{\texttt{$shot=1$ to $R$ }}
	 	\State Use ${\cal U}_{\Phi(\vec{x}_i)}$ to prepare initial feature-map state $\proj{\Phi(\vec{x}_i)}$ 
	 	\State Apply discriminator circuit $W(\vec{\theta})$ to the initial feature-map state .
		\State Apply $|C|$ - outcome measurement $\{M_y\}_{y \in C}$
		\State Record measurement outcome label $y$ by setting $\mbox{r}_{y} \rightarrow \mbox{r}_{y} + 1$
	\EndFor
	\State Construct empirical distribution $ \hat{p}_y(\vec{x}_i) = \mbox{r}_{y} R^{-1}$.  
	\State Evaluate  $\text{Pr} \left( \tilde{m}(\vec{x_i}) \neq y_i | m(\vec{x}) = y_i \right)$ with $\hat{p}_y(\vec{x}_i)$ and $y_i$ 
	\State Add contribution $\text{Pr} \left( \tilde{m}(\vec{x_i}) \neq y_i | m(\vec{x}) = y_i \right)$  to cost function $R_{\textrm{emp}}(\vec{\theta})$.
\EndFor
\State Use optimization routine to propose new $\vec{\theta}$ with information from $R_{\textrm{emp}}(\vec{\theta})$ 
\EndWhile
\State{\Return the final parameter  $\vec{\theta}^*$ and value of the cost function $R_{\textrm{emp}}(\theta^*)$}
\end{algorithmic}
\end{algorithm}

The classification can be applied when the training phase is complete. The optimal parameters are used to decide the correct label for new input data.  Again, the same circuit is applied as in Fig~\ref{fullShortCircuit}, however, this time the parameters are fixed and and the outcomes are combined to determine the label which is reported as output of the classifier. 

\begin{algorithm}[H]
\caption{Quantum variational classification: the classification phase}
\begin{algorithmic}[1]
\State  {\bf Input} An unlabeled sample from the test set $\vec{s} \in S$, optimal parameters $\vec{\theta}^*$ for the discriminator circuit.
\State  {\bf Parameters} Number of measurement shots $R$
\State  Calibrate the quantum Hardware to generate short depth trial circuits.
\State Set the counter $\mbox{r}_{y} = 0$ for every $y \in C$.
	\For{\texttt{$shot=1$ to $R$ }}
	 	\State Use ${\cal U}_{\Phi(\vec{s})}$ to prepare initial feature-map state $\proj{\Phi(\vec{s})}$ 
	 	\State Apply optimal discriminator circuit $W(\vec{\theta^*})$ to the initial feature-map state .
		\State Apply $|C|$ - outcome measurement $\{M_y\}_{y \in C}$
		\State Record measurement outcome label $y$ by setting $\mbox{r}_{y} \rightarrow \mbox{r}_{y} + 1$
	\EndFor
	\State Construct empirical distribution $ \hat{p}_y(\vec{s}) = \mbox{r}_{y} R^{-1}$. 
	\State Set  $\mbox{label} = \mbox{argmax}_{y} \{\hat{p}_y(\vec{s})\}$
\State{\Return $\mbox{label}$}
\end{algorithmic}
\end{algorithm}

\subsubsection*{Quantum kernel estimation}
For the second classification protocol, we restrict ourselves to the binary label case, with $C = \{+1,-1\}$. In this protocol we only use  the quantum computer to estimate the $|T| \times |T|$  kernel matrix $K(\vec{x}_i,\vec{x}_j) = |\braket{\Phi(\vec{x}_i)}{\Phi(\vec{x}_j)}|^2$. For all pairs of points $\vec{x}_i,\vec{x}_j \in T$ in the the training data, we sample the overlap to obtain the matrix entry in the kernel.  This output probability can be estimated from the circuit depicted in Fig.~ \ref{fig:estimate overlaps}.b. by sampling  the output distribution with $R$ shots and only taking the $0^n$ count. After the kernel matrix for the full training data has been constructed we use the conventional (classical) support vector machine classifier. The optimal hyperplane is constructed by solving the dual problem $L_D$ in eqn. (\ref{dualLD}), which is completely specified after we have been given the labels $y_i$ and have estimated the kernel $K(\vec{x}_i,\vec{x}_j)$. The solution of the optimization problem is given in terms of the support vectors $N_S$ for which $\alpha_i > 0$.\\

In the classification phase,  we want to assign a label to a new datum $\vec{s} \in S$ of the test set. For this, the inner product $K(\vec{x}_i,\vec{s})$ between all support vectors $\vec{x}_i \in T$ with $i \in N_S$ and the new datum $\vec{s}$  has to be estimated on the quantum computer. The new label $\tilde{m}(\vec{s})$ for the datum is assigned according to eqn. (\ref{classFeatureMap}).  Since all support vectors are known from the training phase and we have obtained access to the kernel $K(\vec{x}_i,\vec{s})$ from the quantum hardware, the label can be directly computed.

\subsection*{The Relationship of variational quantum classifiers to support vector machines}
The references \cite{vapnik2013nature,burges1998tutorial} provide a detailed introduction to the construction of support vector machines for pattern recognition. Support vector machines are an important tool to construct classifiers for tasks in supervised learning. We will show that the variational circuit classifier bears many similarities to a classical non-linear support vector machine.\\

\subsubsection*{Support vector machines (SVM):} First, let us briefly review the training task of classical, linear support vector machines for data where $C = \{+1,-1\}$, so that  $(\vec{x}_i,y_i)_{i=1,\ldots,t}$ with $ \vec{x}_i \in T \subset \bR^d$, $y_i \in \{+1,-1\}$ that is linearly separable.  Linear separability asks that the set of points can be split in two regions by a hyperplane $({\bf w},b)$, parametrized by a normal vector $ {\bf w} \in \bR^d$ and a bias $b \in \bR$. The points $\vec{x} \in \bR^d$ that lie directly on the hyperplane satisfy the equation
\be
	{\bf w} \circ \vec{x} + b = 0
\ee
expressed in terms of the inner product $\circ$ for vectors in $\bR^d$. The perpendicular distance of the hyperplane to the origin in $\bR^n$ is given by $b\|{\bf w}\|^{-1}$. The data set $\lbrace {\bf x_i}, y_i \rbrace$ is linearly separable by margin $2{||{\bf w}||}^{-1}$ in $\bR^d$ if there exists a vector $\bf w$ and a $b$, such that: \bq\label{sepSVMcontraints}
	&& y_i \left( {\bf w} \circ \vec{x}_i + b \right) \geq 1. \Sp \forall i =1,\ldots,t
\eq
The classification function $\tilde{m}(\vec{x},({\bf w},b))$ that is constructed from such a hyperplane for any new data point $\vec{x} \in \bR^n$ assigns the label according to which side of the hyperplane the new data-point lies by setting
\be\label{SVMclass}
	\tilde{m}(\vec{x},({\bf w},b)) = \mbox{sign}\left( {\bf w} \circ \vec{x} + b\right).
\ee 
The task in constructing a linear support vector machine (SVM) in this scenario is the following. One is looking for a hyperplane that separates the data, with the largest possible distance between the two separated sets. The perpendicular distance between the plane and two points with different labels is called a margin and such points are referred to as `support vectors'. This means that we want to maximize the margin by minimizing $|| {\bf w} ||$, or equivalently $|| {\bf w} ||^2$ subject to the constraints as given in eqn. (\ref{sepSVMcontraints}), for all data points in the training set $T$. The corresponding cost function can be written as:
\be\label{primalLP}
	L_P = \frac{1}{2}\|{\bf w}\|^2  - \sum_{i=1}^t \alpha_i y_i ({\bf w} \circ \vec{x}_i + b) + \sum_{i=1}^t \alpha_i,
\ee 
where $\alpha_i \geq 0$ are Lagrange multipliers chosen to ensure the constraints are satisfied.\\ 

For non-separable datasets, it is possible to introduce non-negative slack variables  $\{\xi_i\}_{i=1,\ldots,t} \in \bR_0^+$ which can be used to soften the constraints for linear separability of $(\vec{x}_i,y_i)$ to 
\bq
y_i\left({\bf w} \circ \vec{x}_i + b \right) &\geq& (1 - \xi_i), \no
\xi_i &\geq& 0.
\eq  
These slack variables are then used to modify the objective function by $1/2 \|{\bf w}\|^2 \raw 1/2 \|{\bf w}\|^2 + C(\sum_i \xi_i)^r + \sum_{i} \mu_i \xi$. When we choose $r \geq 1$ the optimization problem remains convex and a dual can be constructed. In particular, for $r=1$, neither the $\xi_i$ or their Lagrange multipliers $\mu_i$ appear in the dual Lagrangian.\\

It is very helpful to consider the dual of the original primal problem $L_P$ in eqn. (\ref{primalLP}). The primal problem is a convex, quadratic programming problem, for which the Wolfe dual cost function $L_D$ for the Lagrange multipliers can be readily derived by variation with respect to ${\bf w}$ and $b$. The dual optimization problem is
\bq\label{dualLD}
	L_D = \sum_i \alpha_i - \frac{1}{2}\sum_{i,j} \alpha_i \alpha_j y_i y_j \vec{x}_i \circ \vec{x}_j,
\eq
subject to constraints:
\begin{align*} 
0 &\leq \alpha \leq C, & \sum_i \alpha_i y_i &= 0. 
\end{align*} 

The variables of the primal are given in terms of the dual variables by 
\be\label{w_dl}
 \sum_{i} \alpha_i y_i \vec{x}_i  = {\bf w}
\ee 
and the bias $b$ can be computed from the  Karush-Kuhn-Tucker (KKT) conditions when the corresponding Lagrange multiplier  does not vanish. The optimal variables satisfy the KKT conditions and play an important role in the understanding of the SVM. They are given for primal as
\bq
\partial_{{\bf w}_\beta} L_P  &=& {\bf w}_\beta - \sum_i \alpha_i y_i \vec{x}_{i \beta} = 0 \Sp \mbox{for} \Sp \beta = 1,\ldots,d. \\
\partial_{b} L_P 		   &=& - \sum_{i}\alpha_i y_i = 0 \\
					   &&\Sp \alpha_i \geq 0 \\
					   &&\Sp y_i\left(\vec{x}\circ {\bf w} +b\right) - 1\geq 0  \label{comp_slack1} \\
				           &&\Sp \alpha_i\left(y_i({\bf w} \circ \vec{x}_i + b) - 1\right) = 0  \label{comp_slack2}.
\eq
Note that the condition eqn. (\ref{comp_slack2}) ensures that either the optimal $\alpha_i = 0$ or the corresponding constraint eqn. (\ref{comp_slack2}) is tight. This is a property referred to as complementary slackness, and indicates that only the vectors for which the constraint is tight give rise to non-zero $\alpha_i > 0$. These vectors are referred to as the support vectors and we will write $N_S$ for their index set. The classifier in the dual picture is given by substituting ${\bf w}$ from eqn. (\ref{w_dl}) and $b$ into the classifier eqn. (\ref{SVMclass}). The bias $b$ is obtained for any $i \in N_S$ from the equality in eqn. (\ref{comp_slack1}).\\

The method can be generalized to the case when the decision function does depend non-linearly on the data by using a trick from \cite{boser1992training} and introducing a high-dimensional, non-linear feature map. The data is mapped via 
\be
	\Phi : \bR^d \raw \cH
\ee 
from a low dimensional space non-linearly in to a high dimensional Hilbert-space $\cH$. This space is commonly referred to as the feature space. If a suitable feature map has been chosen, it is then possible to apply the SVM classifier for the mapped data in $\cH$, rather than in $\bR^d$.\\

It is important to note that it is in fact not necessary to construct the mapped data $\Phi(\vec{x}_i)$ in $\cH$ explicitly.  Both the training data, as well as the new data to be classified enters only through inner products, in both the optimization problem for training, c.f. eqn. (\ref{dualLD}), as well as in the classifier, eqn. (\ref{SVMclass}). Hence, we can construct the SVM for arbitrarily high dimensional feature maps $(\Phi,\cH)$, if we can efficiently evaluate the inner products $\Phi(\vec{x}_i) \circ \Phi(\vec{x}_j)$ and  $\Phi(\vec{x}_i) \circ \Phi(\vec{s})$, for $\vec{x}_i \in T$ and $\vec{s} \in S$. In particular, if we can find a kernel $K(\vec{x},\vec{y}) = \Phi(\vec{x}) \circ \Phi(\vec{y})$ that satisfies Mercer's condition (which ensures that the kernel is positive semi-definite and can be interpreted as matrix of inner products) \cite{boser1992training,vapnik2013nature}, we can construct a classifier by setting
\be \label{classFeatureMap}
	\tilde{m}(\vec{s}) = \mbox{sign}\left(\sum_{i \in N_S} \alpha_i y_i K(\vec{x}_i,\vec{s}) \; +  b\right).
\ee
Here we only  need to sum over all support vectors $i \in N_S$ for which $\alpha_i > 0$. Moreover, one can replace the inner product in the optimization problem eqn. (\ref{dualLD}) by the kernel. Examples of such kernels that are frequently considered in the classical literature are for instance the polynomial kernel $K(\vec{x},\vec{y}) = (\vec{x} \circ \vec{y} +1)^d$ or even the infinite dimensional Gaussian kernel $K(\vec{x},\vec{y}) = \exp(-1/2\|\vec{x}-\vec{y}\|^2)$. If the feature map is sufficiently powerful, increasingly complex distributions can be classified. In this paper, the feature map is a classical to quantum mapping by a tunable quantum circuit family, that maps $\Phi : \bR^d \raw \cS(\cH_2^{\otimes n})$ in to the state space, or space of density matrices,  of $n$ qubits with $\mbox{dim}\left( \cS(\cH_2^{\otimes n})\right) = 4^n$.  The example of the Gaussian kernel indicates, that the sheer dimension of the Hilbert space on a quantum computer by itself does not provide an advantage, since classically even infinite dimensional spaces are available by for instance using the Gaussian kernel. However,  this hints towards  a potential source of quantum advantage as we may construct states in feature space with hard-to-estimate overlaps.\\

\subsubsection*{Variational circuit classifiers:}  Let us now turn to the case of binary classification based on variational quantum circuits. Recall that in our setting, we first take the data $\vec{x} \in \bR^d$ and map it to a quantum state $\proj{\Phi(\vec{x})} \in \cS(\cH_2^{\otimes n})$ on $n$-qubits, c.f. eqn. (\ref{feature_map_app}). Then we apply a variational circuit $W(\vec{\theta})$ to the initial state that depends on some variational parameters $\vec{\theta}$, c.f. eqn. (\ref{trial_app}). Lastly, for a binary classification task, we measure the resulting state in the canonical $Z$-basis and assign the resulting bit-string $z \in \{0,1\}^n$ to a label based on a predetermined boolean function $f: \{0,1\}^n \rightarrow \{+1,-1\}$. Hence the probability of measuring either label $y \in \{+1,-1\}$ is given by:
\be\label{probabilty_proj}
	p_y = \frac{1}{2}\left(1 + y \bra{\Phi(\vec{x})}W^\dag(\theta)\;{\bf f} \; W(\theta) \ket{\Phi(\vec{x})}\right),
\ee
where we have defined the diagonal operator
\be
	{\bf f} = \sum_{z \in \{0,1\}^n} f(z) \proj{z}.
\ee
In classification tasks we assign c.f. eqn. (\ref{multi_label_decision_rule}), the label with the highest empirical weight of the distribution $p_y$. We ask whether the outcome $+1$ is more likely than $-1$, or vice versa. That is, we ask, whether $p_{+1} > p_{-1} - b$ or whether the converse is true. This of course depends on the sign of the expectation value $\bra{\Phi(\vec{x})}W^\dag(\theta,\varphi)\;{\bf f} \; W(\theta,\varphi) \ket{\Phi(\vec{x})}$ for the data point $\vec{x}$.\\

To understand how this relates to the SVM in greater detail, we need
to choose an orthonormal operator basis, such as for example the Pauli
group 
\be
	\cP_{n} = \avr{X_i,Y_i,Z_i}_{i=1,\ldots,n}.
\ee 
Note that when fixing the phase to $+1$ every element $P_\alpha \in \cP_n$ , with $\alpha = 1,\ldots, 4^n$ of the Pauli-group is an orthogonal reflection $P_\alpha^2 = \1$. Furthermore, Pauli matrices are mutually orthogonal in terms of the trace inner product
\be
	\tr{P_\alpha P_\beta} = \delta_{\alpha,\beta} 2^n.
\ee
This means that both the measurement operator $W^\dag(\theta)\,{\bf f}\,W(\theta)$ in the $W$-rotated frame as well as the state $\proj{\Phi(\vec{x})}$ can be expanded in terms of the operator basis with only real coefficients as
\bq
W^\dag(\theta,\varphi)\;{\bf f} \; W(\theta,\varphi) &=& \frac{1}{2^n}\sum_\alpha w_\alpha(\theta,\varphi) P_\alpha \Sp \mbox{with} \Sp  w_\alpha(\theta,\varphi)  = \tr{ W^\dag(\theta,\varphi)\;{\bf f} \; W(\theta,\varphi)P_\alpha}\no
\proj{\Phi(\vec{x})} &=& \frac{1}{2^n}\sum_\alpha \Phi_\alpha(\vec{x}) P_\alpha \Sp \mbox{with} \Sp   \Phi_\alpha(\vec{x})  = \tr{ \proj{\Phi(\vec{x})}P_\alpha}.
\eq
Note, that the values $w_\alpha(\theta,\varphi)$ as well as $\Phi_\alpha(\vec{x})$ are constrained due to the fact that they originate from a rotated projector and  from a pure state. Since ${\bf f}^2 = \1$, we have that \\ $\tr{\left(W^\dag(\theta,\varphi)\;{\bf f} \; W(\theta,\varphi)\right)^2} = 2^n$. Furthermore, the projector squares to itself so that\\ $\tr{ \proj{\Phi(\vec{x})}^2} = 1$. In particular, this means that the norms of both vectors satisfy \\ $\sum_\alpha \Phi^2_\alpha(\vec{x}) = 2^n$ as well as $\sum_\alpha w^2_\alpha(\theta,\varphi) = 4^{n}$. Since the expectation value of the measured observable is  $\bra{\Phi(\vec{x})}W^\dag(\theta,\varphi) {\bf f} W(\theta,\varphi) \ket{\Phi(\vec{x})} =\tr{ \proj{\Phi(\vec{x})}W^\dag(\theta,\varphi)\;{\bf f} \; W(\theta,\varphi)}$, it can be expressed in terms of the inner product:
\be \label{expansion_SVMQM}
\bra{\Phi(\vec{x})}W^\dag(\theta)\;{\bf f} \; W(\theta) \ket{\Phi(\vec{x})}  =  \frac{1}{2^n}\sum_\alpha w_\alpha(\theta) \Phi_\alpha(\vec{x}).
\ee
Observe that ${\bf f}$ only has eigenvalues ${+1,-1}$, and we have that $\bra{\Phi(\vec{x})}W^\dag(\theta)\;{\bf f} \; W(\theta) \ket{\Phi(\vec{x})} \in [-1,+1]$.  Let us now consider a decision rule, where we assign the label $y \in \{+1,-1\}$ over the label $-y$ with some fixed bias $b \in[-1,+1]$. In that case we demand that  $p_y  >  p_{-y} - yb$. If we substitute eqn. (\ref{probabilty_proj}) and use the expansion in eqn. (\ref{expansion_SVMQM}) we have that the corresponding label is given by the decision function $y = \tilde{m}(\vec{x})$, where
\be
\tilde{m}(\vec{x}) = \mbox{sign}\left(\frac{1}{2^n}\sum_\alpha w_\alpha(\theta)\Phi_\alpha(\vec{x}) + b \right).
\ee
This expression is identical to the conventional SVM classifier, c.f. eqn. (\ref{SVMclass}), after the feature map has been applied. However,  in the experiment we only have access to the probabilities $p_y$ through estimation. Furthermore, the $w_\alpha(\theta)$ are constrained to stem from the observable ${\bf f}$ measured in the rotated frame.\\

This means, that the correct feature space, where a linearly separating hyperplane is constructed is in fact the quantum state space of density matrices, and not the Hilbert space $\cH_2^n$ itself. This is reasonable, since the physical states in  $\cH_2^n$ are only defined up to a global phase $\ket{\psi} \sim e^{i\eta} \ket{\psi}$. The equivalence of states up to a global phase would make it impossible to find a separating hyperplane, since both $\ket{\psi}$ and $-\ket{\psi}$ give rise to the same physical state but can lie on either side of a separating plane.  

\subsection*{Encoding of the data using a suitable feature map}
In the quantum setting, the feature map is an injective encoding of classical information $\vec{x} \in \bR^d$ into a quantum state $\proj{\Phi}$ on an $n$-qubit register. Here $\cH_2 = \bC^2$ is a single qubit Hilbert space, and  $\cS\left( \cH_2^{\otimes n} \right)$  denotes the cone of positive semidefinite density matrices $\rho \geq 0$ with unit trace $\tr{\rho} =1$. This cone is a subset of the $4^n$ dimensional Hilbert space of $\cM_{2^n \times 2^n}(\bC)$ of complex matrices when fitted with the inner product $\tr{A^\dagger B}$ for $A,B \in \cM_{2^n \times 2^n}(\bC)$. The feature map acts as

\begin{align} \label{feature_map_app}
\Phi: \Omega \subset {\bR}^d &\rightarrow \cS\left( \cH_2^{\otimes n} \right), & \Phi :  \vec{x} &\mapsto \proj{\Phi(\vec{x})}.
\end{align}
The action of the map can be understood by a unitary circuit family denoted by ${\cal U}_{\Phi(\vec{x})}$ that is applied to some reference state, e.g. $\ket{0}^n$.  The resulting state is given by $\ket{\Phi(\vec{x})} = {\cal U}_{\Phi(\vec{x})} \ket{0}^n$. The state in the feature space should depend non-linearly on the data. Let us discuss proposals for possible feature maps\\

\paragraph*{Product state feature maps} There are many choices for the feature map $\Phi$. Let us first discuss what would happen if we were to choose a feature map that corresponds to a product input state. We assume a feature map, comprised of single qubit rotations $U(\varphi) \in \text{SU}(2)$ on every qubit on the quantum circuit. The angles for every qubit can be chosen as a non-linear function $\varphi : \vec{x} \raw (0,2 \pi]^2 \times [0, \pi]$ into the space of Euler angles for the individual qubits, so that the full feature map can be implemented as:
\begin{align}
\label{single_feature}
 \vec{x} &\mapsto \ket{\phi_i(\vec{x})} = U(\varphi_i(\vec{x}))\ket{0}, \Sp \mbox{for an individual qubit, so that} \\ 
\Phi : \vec{x} &\mapsto \proj{\Phi(\vec{x})} = \bigotimes_{i=1}^n \proj{\phi_i(x)} \Sp \mbox{for the full qubit state.}
\end{align}
One example for such an implementation is the unitary implementation of the feature map used in the context of the classical classifiers by Stoudenmire and Schwab \cite{stoudenmire2016supervised}  based on tensor networks. There each qubit encodes a  single component $x_i$ of $\vec{x} \in [0,1]^n$ so that $n$ qubits are used. The resulting state that that is prepared is then
\be
\bigotimes_{i=1}^n \proj{\phi_i(x)} = \frac{1}{2^n} \bigotimes_{j = 1}^n \left( \sum_{\alpha_j} \Phi_j^{\alpha_j} \left( \theta_j(\vec{x}) \right) P_{\alpha_j} \right),
\ee
when expanded in terms of the Pauli-matrix basis where $\Phi^\alpha_i(\theta_i(\vec{x})) =  \bra{\phi_i(x)} P_{\alpha_i} \ket{\phi_i(x)}$ for all $i=1, \ldots n$.  and $P_{\alpha_i} \in \{\1,X_i,Z_i,Y_i\}$. The corresponding decision function can be constructed as in eqn. (\ref{classFeatureMap}), where the kernel $K(\vec{x},\vec{y}) = \prod_{i=1}^n |\braket{\phi_i(\vec{x})}{\phi_i(\vec{y})}|^2$ is replaced by the inner product between the resulting product states. These can be evaluated with resources scaling linearly in the number of qubits, so that no quantum advantage can be expected in this setting.\\

\subsubsection*{Non-trivial feature map with entanglement} 
There are many choices of feature maps, that do not suffer from the malaise of the aforementioned product state feature maps. To obtain an quantum advantage we would like these maps to give rise to a kernel 
$K(\vec{x},\vec{y}) = |\braket{\Phi(\vec{x})}{\Phi(\vec{y})}|^2$ that is computationally hard to estimate up to an additive polynomially small error by classical means. Otherwise the map is immediately amenable to classical analysis and we are guaranteed to have lost any conceivable quantum advantage. 

\begin{figure}[h]
\begin{center}
\includegraphics[scale=0.7]{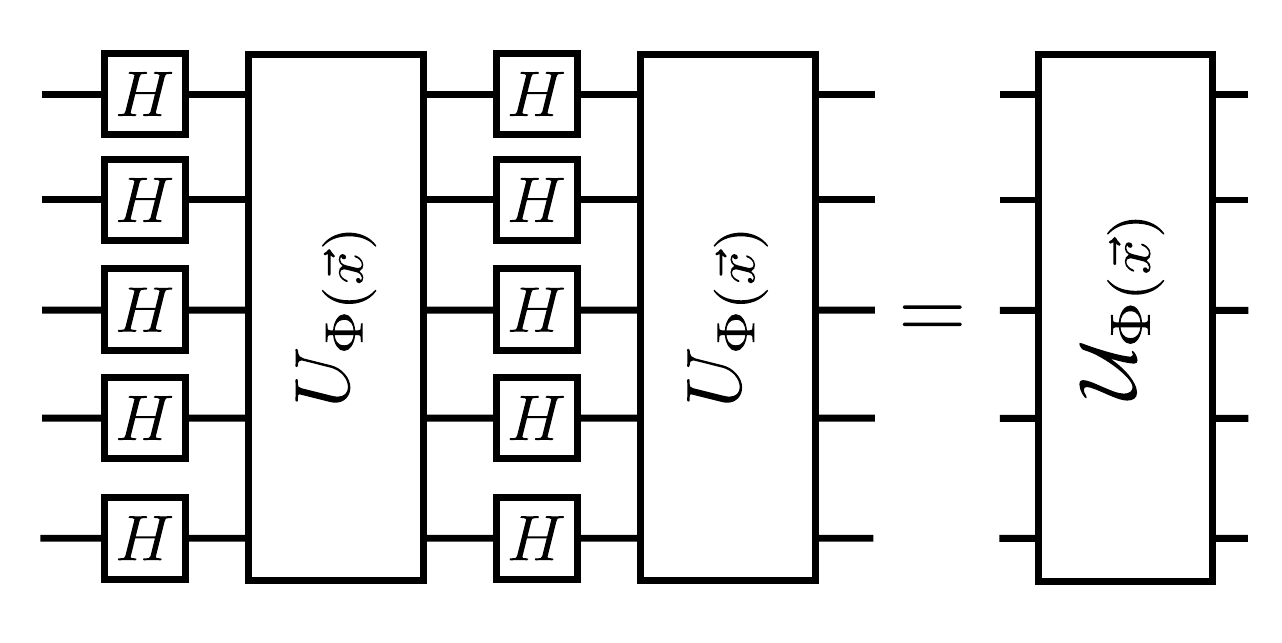}
\caption{ \label{fig:feature_map1} A circuit representation of the feature map family we consider here.  We first apply a series of Hadamard gates before applying the diagonal phase gate component. Then we apply a second layer of Hadamard gates, followed by the same diagonal phase gate. This encodes both the actual function value of the phase $\Phi_{\vec{x}}(z)$  as well as the value of the $\bZ^N_2$ Fourier transform $\hat{\Phi}_{\vec{x}}(S)$  for every basis element.}
\end{center}
\end{figure}

Let us therefore turn to a family of feature maps, c.f. Fig~\ref{fig:feature_map1} for which we conjecture that it is hard to estimate the overlap $|\braket{\Phi(\vec{x})}{\Phi(\vec{y})}|^2$ on a classical computer. 
We define the family of feature map circuit as follows
\be \label{feature_map_circ_app}
\ket{\Phi(\vec{x})}  \; = \;  U_{\Phi(\vec{x})} \; H^{\otimes n}   \; U_{\Phi(\vec{x})}  H^{\otimes n} \ket{0}^{\otimes n} \Sp \mbox{where,} \Sp
U_{\Phi(\vec{x})}  \; = \; \exp\left( i \sum_{S \subseteq [n]} \phi_S(\vec{x}) \prod_{i \in S} Z_i\right). 
\ee
where now the $2^n$ possible coefficients $\phi_S(\vec{x}) \in \bR$ are non-linear functions of the input data $\vec{x} \in \bR^n$. It is convenient to use maps with low-degree expansions, i.e. $|S| \leq d$.  
Any such map can be efficiently implemented. In the experiments reported in this paper we have restricted to $d=2$. So we only consider Ising type interactions in the unitaries  $U_{\Phi(\vec{x})}$. In particular we choose these interactions as the ones that are present in the actual connectivity graph of the superconducting chip $G = (E,V)$. This ensures that the feature map can be generated from a short depth circuit. The resulting unitary can then be generated from one-  and two- qubit gates of the form
\be\label{feature_map_entangler}
	U_{\phi_{\{l,m\}}(\vec{x})}   =  \exp\left( i \phi_{\{k,l\}}(\vec{x}) Z_k Z_l\right) \Sp \mbox{and} \Sp  U_{\phi_{\{k\}}(\vec{x})}  = \exp\left(i \phi_{\{k\}}(\vec{x}) Z_k \right),
\ee
which leaves $|V| + |E|$, real parameters to encode the data. In particular, we know that we have at least $|V| = n$ real numbers to encode the data. Furthermore, depending on the connectivity of the interactions, we have $|E| \leq n(n-1)/2$ further parameters that can be used to encode more data or nonlinear relations of the initial data points.

This feature map encodes both the actual function $\Phi_{\vec{x}}(z)$ of the diagonal phases, as well as the corresponding Fourier-Walsh transform $\hat{\Phi}_{\vec{x}}(p)$ at $z,p \in \{0,1\}^n$ 
\be \label{classical_diagonal}
 \Phi_{\vec{x}}(z) = \exp\left( i \sum_{S \subseteq [n]} \phi_S(\vec{x}) \prod_{i \in S}(-1)^{z_i}\right) \Sp \mbox{and} \Sp \hat{\Phi}_{\vec{x}}(p) = \frac{1}{2^n} \sum_{z \in \{0,1\}^n} \Phi_{\vec{x}}(z) (-1)^{p \circ z},
\ee
for every basis element respectively. The resulting state, after the datum $\vec{x} \in \bR^d$ has been mapped to the feature space, is given by
\be
\ket{\Phi(\vec{x})} = \sum_{p \in \{0,1\}^N}  \Phi_{\vec{x}}(p)  \hat{\Phi}_{\vec{x}}(p) \ket{p}.
\ee
We conjecture that it is hard to estimate this kernel up to an additive polynomially small error by classical means. The intuition for the conjecture stems from a connection of the feature map to a particular circuit family for the hidden shift problem for boolean functions \cite{van2006quantum}. The feature map is similar to a quantum algorithm for estimating the hidden shift in boolean bent functions, c.f. Ref.  \cite{rotteler2010quantum}, algorithm ${\cal A}_1$ in Theorem 6. The circuit $\cU_\Phi$ we consider is indeed very similar to the one discussed in \cite{rotteler2010quantum}. Let us make a minor modification, and ask the two diagonal layers in $\tilde{\cU}_\Phi = U_{\Phi^2}H^{\otimes n}U_{\Phi^1}H^{\otimes n}$ to differ. We choose the phase gate layers so that $U_{\Phi^1}$ and $U_{\Phi^2}$ encode a shifted bent-function and the function's dual, then the circuit ${\cal A}_1$ is given by $H^{\otimes n} \tilde{\cU}_\Phi$.  If the interaction graph $G$ is bi-partite, it is possible to encode Maiorana-McFarland bent-functions for $d = 2$ by choosing the corresponding $\phi_{\{k,l\}}(\vec{x}), \phi_{k}(\vec{x})$ to be either $\pi$ or $0$. In \cite{rotteler2010quantum}, the diagonal layer in ${\cal A}_1$, are queries to an oracle encoding the the shifted bent-function and it's dual. It can be shown that with respect to this oracle there is an exponential separation in query complexity over classical query algorithms. The final circuit we implement to estimate the overlap $|\braket{\Phi(\vec{x})}{\Phi(\vec{y})}|^2$ is still larger. This circuit has four layers of Hadamard gates and 3 diagonal layers. If the conjecture could be proven, it would establish a valuable step in rigorously establishing a quantum advantage on near-term devices. If, however, it ultimately turns out that this family can also be evaluated classically, we would need to improve the complexity of this circuit family. \\

A good example of a circuit family that entails a hardness result, yet does not provide an advantage in our setting is the following:  One could consider the case, where only a single layer of Hadamard gates and a single diagonal unitary is applied. Such a feature map is directly related to the circuit family introduced in \cite{bremner2016average}. Indeed the resulting kernel $K(\vec{x},\vec{y}) = |\bra{0^n} H^{\otimes n} U_{\Phi(\vec{y})} U_{\Phi(\vec{x})} H^{\otimes n} \ket{0^n} |^2$ corresponds to the output probability of the considered IQP circuits. In fact it is known that if general real values $\phi_S(\vec{x})$ are allowed that $d = 2$ suffices to encode $\#P$-hard problems in the output probability \cite{goldberg2017complexity}. This hardness result, only applies to the case where the output probability can be approximated up to a multiplicative error. A noisy quantum computer, however, is also not able to provide a multiplicative error estimate to this quantity. Nevertheless, Bremner et al. show, contingent on additional complexity theoretic conjectures,  that it is hard for a classical sampler to produce samples from the output distribution of the IQP circuit. Does this result imply some form of hardness for the feature map version of this circuit? The answer is unfortunately no. The kernel can be estimated up to an additive error of $\epsilon$ by drawing $R = \epsilon^{-2}$ samples from the uniform distribution over $n$ classical bits and averaging $\tilde{\Phi}_{\vec{x} - \vec{y}}(z) =  \Phi_{\vec{x}}(z) \overline{\Phi}_{\vec{y}}(z) $, c.f. eqn. (\ref{classical_diagonal}) 
\be
 |\bra{0^n} H^{\otimes n} U_{\Phi(\vec{y})} U_{\Phi(\vec{x})} H^{\otimes n} \ket{0^n} |^2 =  \left | \frac{1}{2^n}\sum_{z \in \{0,1\}^n} \tilde{\Phi}_{\vec{x} - \vec{y}}(z) \right |^2.
\ee
Since we have that the variance of the  random variable is bounded by $1$ since $|\Phi_{\vec{x}}(z)|^2  = 1$ , we get an additive error that scales as $\cO(\epsilon)$. This means for a single layer, the kernel can be estimated classically.

\subsection*{Quantum variational classification}
\begin{figure}[h]
\begin{center}
\includegraphics[scale=0.4]{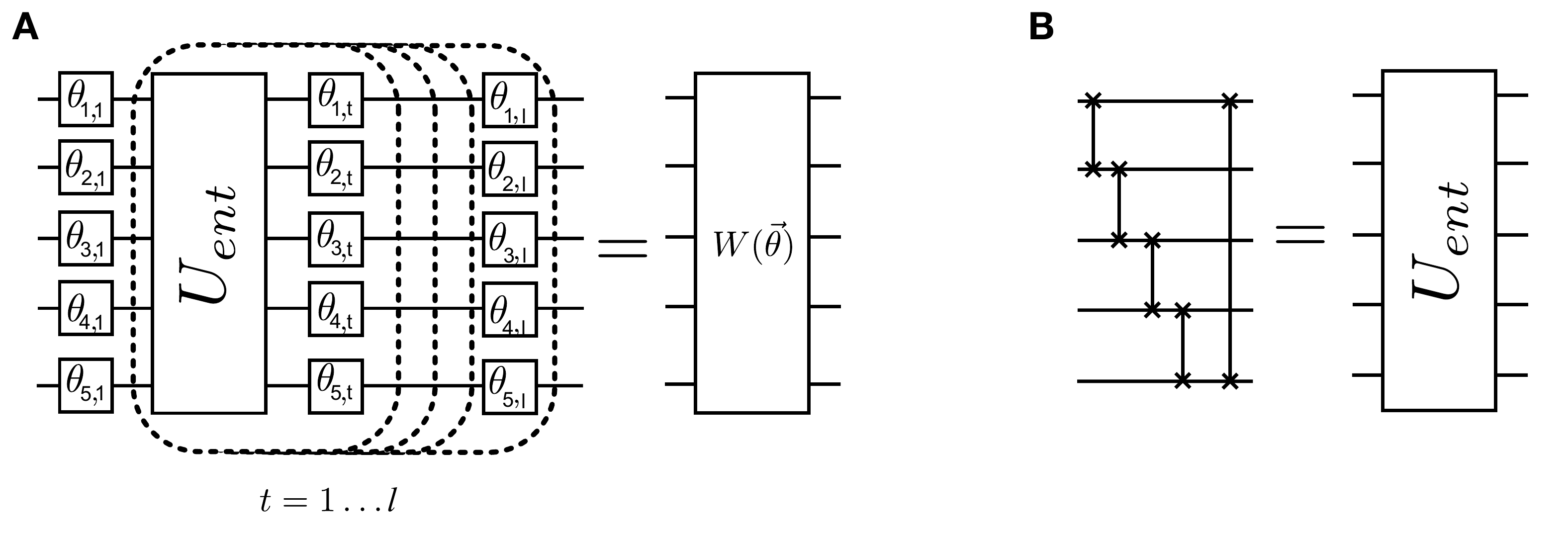}
\caption{\label{circuit} (a) Circuit representation of short depth quantum circuit to define the separating hyperplane. The single qubit rotations $U(\theta_{i,t}) \in \text{SU}(2)$ are depicted by single line boxes parametrized by the angles $\theta_i$, while the entangling operation $U_{ent}$ is determined by the interaction graph of the superconducting chip. (b) Depiction of entangling gate as a product of  $\textsf{CZ}_{i,i+1}$ for $i = 1,\ldots,5$ gates following the interaction graph of a circle $G = C_5$. }
\end{center}
\end{figure}

Following the structure of the feature map circuit, we construct the classifier part of the variational algorithm by appending layers of single-qubit unitaries  and entangling gates Fig.~\ref{circuit}.a. Each subsequent layer, or depth, contains an additional set of entanglers across all the qubits used for the algorithm. We use a coherently controllable quantum mechanical system, such as for example the superconducting chip with $n$ transmon qubits to prepare a short depth quantum circuit $W(\vec{\theta})$. In the experiment here, comprising $n=2$ qubits, one controlled-phase gate is added per depth. The single-qubit unitaries used in the classifier are limited to $Y$ and $Z$ rotations to simplify the number of parameters to be handled by the classical optimizer. Our use of controlled-phase, rather than CNOT, gates for the entanglers is justified by our aim at increased generality in our software. Using controlled-phase gates does not require to particularize this part of the algorithm for different systems topologies. A specific entangling map for a given device can then be used by our compiler to translate each controlled-phase gate into the CNOTs available in our system.  

The general circuit is comprised of the following sequence of single qubit and multi-qubit gates:
\be  \label{trial_app}
W(\vec{\theta}) = U_{\textrm{loc}}^{(l)}(\theta_l) \;U_{\textrm{ent}} \ldots U_{\textrm{loc}}^{(2)}(\theta_2) \; U_{\textrm{ent}}\; U_{\textrm{loc}}^{(1)}(\theta_1). 
\ee
We apply a circuit of $l$ repeated entanglers as depicted in Fig \ref{circuit}.b and interleave them with layers comprised of local single qubit rotations: 
\be
U_{\textrm{{loc}}}^{(t)}(\theta_t) = \otimes_{m=1}^n U(\theta_{m,t}) \Sp \mbox{and} \Sp U(\theta_{m,t}) = e^{i \frac{1}{2}{\theta^z_{m,t}} Z_m}e^{i \frac{1}{2}{\theta^y_{m,t}} Y_m},
\ee 
parametrized by $\theta_t \in \bR^{2n}$ and $\theta_{i,t} \in \bR^{2}$. In principle, there exist multiple choices for the entangling unitaries $U_{\textrm{ent}}$ \cite{kandala2017hardware,farhi2017quantum}. For the feature map that we consider, however, we use the entangler that is comprised of products of control phase gates $\textsf{CZ}(i,j)$ between qubits $i$ and $j$. The entangling interactions follow the interaction graph that $G = (E,V)$ that was used to generate the feature map in eqn. (\ref{feature_map_entangler}).
\be
U_{\textrm{ent}}= \prod_{(i,j) \in E} \textsf{CZ}(i,j),
\ee  
This short-depth circuit can generate any unitary if sufficiently many layers $d$ are applied. The circuit Fig.~\ref{circuit}.a can be understood as a bang-bang controlled evolution of an Ising model Hamiltonian $H_{0} = \sum_{(ij) \in E} J_{ij} Z_i Z_j$,c.f.  Fig.~\ref{circuit}.b, interspersed with single qubit control pulses in $SU(2)$ on every qubit. It is known that this set of drift steps together with all single control pulses  are universal, so we have that any state can be prepared this way with sufficient circuit depth \cite{d2007introduction}. For a general unitary gate sequence the entangling unitary has to be effectively generated from cross resonance gates, by applying single qubit control pulses.

\subsection*{Choosing the cost-function for the circuit optimization}
The central goal is to find an optimal classifying circuit $W(\vec{\theta})$, c..f. eqn. (\ref{trial_app}) that separates the data sets with different labels. Since we can re-run the same classifying multiple times ($R$ shots), we may consider a `winner takes all' scenario, where we assign the label according to the outcome with the largest probability. We choose a cost function, so that the optimization procedure minimizes the probability of assigning the wrong label after having constructed the distribution after $R$ shots.\\

There are multiple ways of performing a multi-label classification. We only need to modify the final measurement  $M$, to correspond to multiple partitions. This can be achieved by multiple strategies. For example one could choose to measure again in the computational basis, i.e. the basis in which Pauli $Z$ are diagonal and then constructing classical labels form the measured samples, such as a labeling the outcome $z \in \{0,1\}^n$ according to a function $f :\{0,1\}^n \raw \{1,\ldots,c\}$.  The resulting $\{M_y\}_{y=1,\ldots,c}$ is therefore diagonal in the computational basis. Alternatively one could construct a commuting measurement akin to the syndrome check measurement for quantum stabilizers. For this approach we choose a set $\{g_i\}_{i=1 \ldots \lceil{\log_2(c)}\rceil}$ of Pauli matrices $g_i \in \cP_N$ that are commuting $[g_i,g_j] = 0$. The resulting measurement that would need to be performed is similar to that of an error correcting scheme. The measurement operators are given by $M_y = 2^{-1}\left(1-\prod_{i=1}^{\lceil{\log_2(c)}\rceil} g_i^{y^i}\right)$. Here $y^i$ denotes the $i$'th bit in the binary representation of $y$. In either case, the decision rule that assigns the labels can be written as
\begin{align} \label{multi_label_decision_rule}
\tilde{m}_{|T}(\vec{x}) = \text{arg}\max_{c'}  \bra{\Phi(\vec{x})} W(\vec{\theta})^\dag M_{c'} W(\vec{\theta}) \ket{\Phi(\vec{x})}\,.
\end{align}
This corresponds to taking $R$ shots in order to estimate the largest outcome probability from the outcome statistics of the measurement $M_y$ for $y = 1,\ldots,c$. Labelling $T_c$ the subset of samples $T$ labelled with $c$, the overall expected misclassification rate is given by:
\begin{align}
P_{\textrm{err}} &= \frac{1}{|T|} \bigg( \sum_c \sum_{s \in T_c} \text{Pr} \left(\tilde{m}_{|T}(s) \neq c  | s \in T_c \right) \bigg)\,.
\label{Eq:Perr}
\end{align}
The error probability of misclassifying an individual datum is given by $\text{Pr} \left(\tilde{m}_{|T}(s) \neq c  | s \in T_c \right)$. This error probability is now used to define the empirical risk function $R_{\textrm{emp}}(\vec{\theta}) = P_{\textrm{err}}$. We now discuss of how to find suitable ways of evaluating the cost function for this classification scheme. 

\subsection*{Binary label classification}
Assume the programmer classifies into labels $y \in \lbrace -1, 1 \rbrace$ by taking $R$ shots for a single datapoint. She obtains an empirical estimates of probability of the datum being labeled by a label $y$
\begin{align*}
\hat{p}_y &= \frac{r_y}{R} \,.
\end{align*}
After $R = r_y + r_{-y}$ shots and a prior bias $b$, she misclassifies into a label $y$ if
\begin{align*} 
\hat{p}_y < \hat{p}_{-y} + yb   \rightarrow r_y &< r_{-y} + ybR \rightarrow  r_y < \left\lceil \frac{1 + yb}{2} R \right\rceil
\end{align*}
The probability of her misclassifying a $y$ sample according to the argmax rule is hence estimated by
\begin{align*} 
\text{Pr} \left(\tilde{m}_{|T}(s) \neq y | s \in T_y \right) &=\text{Pr}\left(r_y < r_{-y} +  yb \right) =  \sum_{j=0}^{\lceil \frac{1+yb}{2} R \rceil} {R \choose j} p_y^j (1-p_y)^{R-j}\,.
\end{align*}
Assuming large $R$, computing this exactly may be difficult. Setting $Rp_y = a, Rp_y(1-p_y) =\beta^2$ and $\lceil \left( \frac{1+yb}{2} \right) R \rceil  =\gamma$, we can approximate the binomial CDF as an error function: 
\begin{align*}
\text{Pr} \left(\tilde{m}_{|T}(s) \neq y | s \in T_y \right)   &= \sum_{j=0}^{\lceil \left( \frac{1+y b}{2} \right) R\rceil } {R\choose j} p_c^{R-j} (1-p_c)^{j} \approx \int_{-\infty}^{\gamma} dx \frac{1}{\sqrt{2 \pi}\beta} {\exp \left(- \frac{1}{2}\left(\frac{x-a}{\beta}\right)^2 \right)} \\
&= \frac{1}{\sqrt{\pi}} \int_{-\infty}^{\frac{\gamma-a}{\sqrt{2}\beta}} dz \; e^{-z^2} =  \frac{1}{2}  \mbox{erf} \left( \frac{\gamma-a}{\sqrt{2}\beta} \right) + \frac{1}{2} \\
&= \frac{1}{2} \mbox{erf} \left(\sqrt{R} \frac{\frac{1 + y b}{2}-p_y}{\sqrt{2 (1-p_y)p_y}} \right) + \frac{1}{2}\,.
\end{align*}
\begin{figure}[h]
\begin{center}
\includegraphics[scale=0.5]{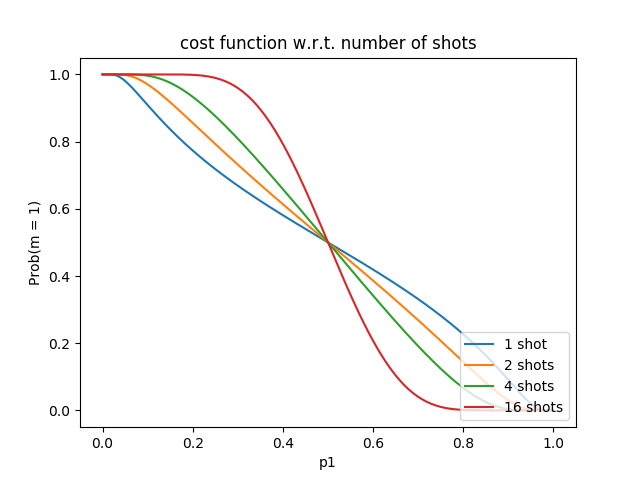}
\caption{\label{Fig_err}
Single shot to multi-shot decision rule for classification. The contribution to the cost-function interpolates from linear to logistic-normal CDF (approximately sigmoid). In the experiment the data was sampled with $R = \cO(10^3)$, although the cost-function was evaluated with only $\tilde{R} = \cO(10^2)$ to provide a smoother function to the optimization routine.}
\end{center}
\end{figure}
See Fig.~\ref{Fig_err}. The error function can be consequently approximated with a sigmoid
\begin{align*}
\text{sig}(x) = \frac{1}{1 + \exp(-x)} \approx \frac{1}{2} \left( \mbox{erf}(x) + 1 \right), \\
\end{align*}
which gives
\begin{align*}
\text{Pr} \left(\tilde{m}_{|T}(s) \neq y | s \in T_y   \right) &\approx  \text{sig}\left(\sqrt{R} \frac{\frac{1 + yb}{2} -p_y}{\sqrt{2 (1-p_y)p_y}}\right) \,.
\end{align*}
as an estimate for misclassifying a sample $s$. The cost function to optimize is then given by using this in eqn.~(\ref{Eq:Perr}).

\subsubsection*{Multi label classification}
For multiple labels, one tries to optimize
\begin{align*}
P_{err} &=  \frac{1}{|T|} \sum_c \sum_{s \in T_c} \text{Pr} \left( \tilde{m}_{|T}(s) \neq c | s \in T_c \right)\,,
\end{align*}
where:
\[
\text{Pr} \left( \tilde{m}_{|T}(s) \neq c | s \in T_c \right) = \text{Pr} \left(n_c < \max_{c'}\left( \lbrace n_{c'} \rbrace_{c'/c} \right) \right)\,.
\]
We consider the case of three labels. For $R$ samples with frequencies $\lbrace n_0, n_1, n_{2} \rbrace$, drawn independently from the output probability distribution, the probability of misclassifying a sample $s \in T_0$ by argmax is given by
\begin{align*}
\text{Pr} \left( \tilde{m}_{|T}(s) \neq 0 | s \in T_0 \right) = \text{Pr} \bigg(n_0 < \max(n_1, n_2)\bigg) = \text{Pr}\bigg(n_0 <  \left\lceil \frac{N + |n_1 - n_2|}{3} \right\rceil \bigg)\,,
\end{align*}
where the last inequality is derived as follows
\begin{align*}
2n_0 & < 2 \max(n_1, n_2) =  |n_1 - n_2| + n_1 + n_2 =  |n_1 - n_2| + N - n_0\,.
\end{align*}
Hence setting $\gamma = \frac{ N + |n_1 - n_2|}{3}$, it follows that 
\begin{align*} 
\text{Pr} \left( \tilde{m}_{|T}(s) \neq 0 | s \in T_0 \right)  &= \sum_{k= 0}^{k=\gamma}{R \choose k} p_0^k (1-p_0)^{N-k} \approx  \text{sig}\left( \frac{\gamma-Np_0}{\sqrt{2 N(1-p_0) p_0}}\right)\,. 
\end{align*}
This however still depends on $n_1, n_2$, which can't be simply eliminated. Additionally, for a general $k$-label case, there is no simple analytic solution for $\gamma$. For this reason, we therefore try to estimate the above probability by simply taking $\gamma = \max_{c'}\left( \lbrace n_{c'} \rbrace_{c'/c} \right)$. So for $k$-label case, the cost function terms are given by 
\begin{align*} \text{Pr} \left( \tilde{m}_{|T}(s) \neq c | s \in T_c \right) \approx \text{sig}\left( \sqrt{R}\frac{\max_{c'}\left( \lbrace n_{c'} \rbrace_{c'/c} \right)-n_c}{\sqrt{2 (N-n_c) n_c}}\right)\,. 
\end{align*}

\subsection*{Quantum kernel estimation}
For the second classification method we only use the quantum computer to estimate the kernel $K(\vec{x}_i,\vec{x}_j) = |\braket{\Phi(\vec{x}_i)}{\Phi(\vec{x}_j)}|^2$ for all the labeled training data $\vec{x}_j \in T$. Then we use the classical optimization problem as outlined again in eqn. (\ref{dualLD}) to obtain the optimal Lagrange multipliers $\alpha_i$ and support vectors $N_S$ can be obtained. From this the classifier can be constructed, c.f .eqn. (\ref{classFeatureMap}). To apply the classifier to a new datum $\vec{s} \in S$ the kernel  $K(\vec{x}_i,\vec{s})$ between $\vec{s}$ and the support vectors in $i \in N_S$ has to be estimated. We discuss two methods to estimate this overlap for our setting.

\begin{figure}[h]
\begin{center}
\includegraphics[scale=1.6]{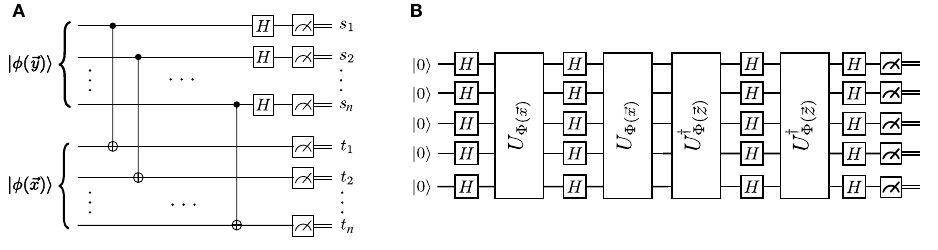}
\caption{ \label{fig:estimate overlaps} a) Estimating the expectation value of the $\textsf{SWAP}$- matrix as derived in \cite{cincio2018learning}. This circuit diagonalizes the $\textsf{SWAP}$ - gate, when it is treated as a  hermitian observables with eigenvalues $\pm 1$. Averaging the eigenvalues, c.f. eqn. (\ref{swap_ev}), with samples from the output distribution constructs an estimator for  $|\braket{\Phi(\vec{x})}{\Phi(\vec{y})}|^2$. b) The ciruit estimates the fidelity between two states in feature space directly by first applying the unitary ${\cal U}_{\Phi(\vec{x})}$ followed by the inverse  ${\cal U}^\dagger_{\Phi(\vec{y})}$ and measuring all bits at the output of the circuit. The frequency $\nu_{0,\ldots,0} = |\braket{\Phi(\vec{x})}{\Phi(\vec{y})}|^2$  of the all the zero outcome precisely corresponds to the desired state overlap.}
\end{center}
\end{figure}

The usual method of estimating the fidelity between two states is by using the swap test \cite{buhrman2001quantum}. This circuit, however, is not a short depth circuit on a quantum computing architecture with geometrically local gates. It would require a sequence of controlled $\textsf{SWAP}$, also known as Fredkin gates, all conditioned on the state of the same ancilla qubit. A very nice protocol was recently developed in \cite{cincio2018learning}. The authors have learned multiple ways of optimizing the conventional swap test. If only the value of the fidelity is needed, as is the case for our algorithm, the authors propose, c.f. \cite{cincio2018learning} section III.C,  a circuit that is constant depth if pairs of $\textsf{CNOT}$ gates can be executed in parallel. This circuit, c.f. Fig~\ref{fig:estimate overlaps}.a, evaluates the expectation value $\bra{\psi}\bra{\phi} \textsf{SWAP} \ket{\psi}\ket{\phi}$ directly. The action of the algorithm can be understood as follows: 

The $\textsf{SWAP}$ gate is both a unitary gate and a hermitian observable $\textsf{SWAP}^\dagger = \textsf{SWAP}$ with eigenvalues $\pm 1$. The expectation value on two product states as we said given by \\ $\bra{\psi}\bra{\phi} \textsf{SWAP} \ket{\psi}\ket{\phi} = |\braket{\phi}{\psi}|^2$. Now the gate can be decomposed in to a product of two qubit swap gates $\textsf{SWAP} = \prod_{k=1}^n \textsf{SWAP}_{s_k t_k}$ that act all in parallel. To evaluate the expectation value one has to diagonalize the full gate. This is achieved by diagonalizing the individual two-qubit swap gates by observing that $\textsf{SWAP}_{ij} = \textsf{CNOT}_{i\raw j} \textsf{CNOT}_{j\raw i} \textsf{CNOT}_{i\raw j}$. Furthermore using the circuit identity $\textsf{CNOT}_{j\raw i} = H_j \textsf{CZ}_{j i} H_j$, we see that $\textsf{SWAP}_{ij}$ is diagonalized by $\textsf{CNOT}_{j\raw i} H_j$ and has eigenvalue $(-1)^{x_ix_j}$. For the full circuit Fig~\ref{fig:estimate overlaps}.a one first applies a transversal set of $\textsf{CNOT}$ gates across both registers followed by a single layer of Hadamard gates $H$ on the top register. Then the output is sampled and the average of the boolean function
\be \label{swap_ev}
f(s,t) = (-1)^{(s_1t_1 + \ldots s_n t_n)}
\ee
 is reported. The output bits on the top register are labeled by $s \in \{0,1\}^n$, while $t \in \{0,1\}^n$ is the output string on the lower register.\\
 
This method works for arbitrary input states $\ket{\psi},\ket{\phi}$. Our states, however, are structured and are all generated by the unitary eqn. (\ref{feature_map_circ_app}) as shown in Fig ~\ref{fig:feature_map1}. Writing out the kernel explicitly $K(\vec{y},\vec{x}) = |\braket{\Phi(\vec{y})}{\Phi(\vec{x})}|^2 = |\bra{0^n}{\cal U}^\dagger_{\Phi(\vec{y})}{\cal U}_{\Phi(\vec{x})}\ket{0^n}|^2$ gives the indication of how to measure  it. Simply apply the circuit ${\cal U}^\dagger_{\Phi(\vec{y})}{\cal U}_{\Phi(\vec{x})}$ to the state $\ket{0^n}$, c.f. Fig~\ref{fig:estimate overlaps}.b. Then sample $R$ times the resulting state ${\cal U}^\dagger_{\Phi(\vec{y})}{\cal U}_{\Phi(\vec{x})}\ket{0^n}$ in the $Z$ basis. Record the number of all observed zero $(0,\ldots,0)$ bit-strings and divide by the total number of shots $R$. The frequency $\nu_{(0,\ldots,0)} = \#\{(0,\ldots,0)\}R^{-1}$ then gives an estimator for $K(\vec{y},\vec{x})$ up to a sampling error $\tilde{\epsilon} = \cO(R^{-1/2})$. A rough estimate for the operator norm $\|\cdot\|$ of the deviation of the resulting estimator $\hat{K}$ from the true kernel matrix $K$  is  given by $\| K  - \hat{K}\| \leq \| K  - \hat{K}\|_{F}$. Here  $\|A\|_{F} = \sqrt{\sum_{ij}|A_{ij}}|^2$ denotes the Frobenius norm of the matrix $A$. A crude bound can then be obtained from the largest sampling error $\tilde{\epsilon}$ of all matrix entries and setting $\| K  - \hat{K} \|_{F} \leq \tilde{\epsilon} |T|$, since both matrices of the training set $T$ are of size $|T| \times |T|$. Hence to ensure a maximum deviation of $\epsilon$ with high probability a number of $R = \cO(\epsilon^{-2} |T|^2)$ shots have to be drawn for each matrix entry.  Due to the symmetry of the $K$ matrix and the trivial diagonals,  $|T|(|T|-1)2^{-1}$ matrix entries have to be estimated. Thus the full sampling complexity is expected to scale as $\cO(\epsilon^{-2} |T|^4)$. A more careful analysis of the statistical error could be carried out by using one of the matrix-concentration results \cite{tropp2015introduction}.\\ 

Note that the optimization problem, eqn.  (\ref{dualLD}) is only concave, when the matrix $K \geq 0$ is positive semi-definite. It can happen, that the shot noise and other errors in the experiment lead to a $\hat{K}$ that is no longer positive semi-definite. We have indeed observed this multiple times in the experiment. A possible way of dealing with this problem is a method developed in \cite{smolin2012efficient}, where an optimization problem is solved to find the closest positive semi-definite $K$-matrix in trace norm to $\hat{K}$ consistent with the constraint. In our experiments however, we have found this not to be necessary and the performance has been almost optimal without performing this method.  

\subsection*{Device parameters}
Our device is fabricated on a 720-$\mu$m-thick Si substrate. A single optical lithography step is used to define all CPW structures and the qubit capacitors with Nb. The Josephson junctions are patterned via electron beam lithography and made by double-angle deposition of Al.

The dispersive readout signals are amplified by Josephson Parametric Converters \cite{Bergeal2010} (JPC). Both the quantum processor and the JPC amplifiers are thermally anchored to the mixing chamber plate of a dilution refrigerator.

The two qubit fundamental transition frequencies are $\omega_i/2\pi = \{ 5.2760(4), 5.2122(3) \}$ GHz, with anharmonicities $\Delta_i/2\pi = \{ -330.3,    -331.9\}$ MHz, where $i \in \{0,1\}$. The readout resonator frequencies used are $\omega_{Ri}/2\pi=\{6.530553,   6.481651\}$ GHz, while the CPW bus resonator connecting $Q_0$ and $Q_1$ was unmeasured and designed to be 7.0 GHz. The dispersive shifts and line widths of the readout resonators are measured to be $2\chi_i/2\pi = \{-1.06, -1.02 \}$ MHz and $\kappa_i/2\pi = \{661,681  \}$ kHz, respectively.

The two qubit lifetimes and coherences were measured intermittently throughout our experiments. The observed mean values were $T_{1(i)} = \{55, 38 \}$, $T^*_{2(i)} = \{16, 17 \}$, $T^{\rm{echo}}_{2(i)} = \{43, 46 \}$ $\mu$s with $i \in \{0,1\}$

\subsection*{Gate characterization}
Our experiments use calibrated $X-$rotations ($X_\pi$ and $X_{\pi/2}$)  as single-qubit unitary primitives. $Y-$rotations are attained by appropiate adjustment of the pulse phases, whereas $Z-$rotations are implemented via frame changes in software \cite{mckay2017efficient}. A time buffer is added at the end of each physical pulse to mitigate effects from reflections and evanescent waves in the cryogenic control lines and components.

We use two sets of gate times in order to perform the Richardson extrapolation of the noise in our system \cite{temme2017error}. For the first set of gate times we use 83 ns for single-qubit gates and 333 ns for each cross-resonance pulse. The buffer after each physical pulse is 6.5 ns. The single-qubit gates are gaussian shaped pulses with $\sigma=20.75$ ns. The cross-resonance gates are flat with turn-on and -off gaussian shapes of $\sigma=10$ ns. Our implementation of a CNOT has a duration of two single-qubit pulses and two cross-resonance pulses, giving a total of 858 ns for the first set of gate times, including buffers. For our second set of gate times we use the times of the first set but stretched by a factor of 1.5, including the pulses $\sigma$s and the buffers. This gives a total CNOT time of 1.287 $\mu$s and single-qubit gates of $\sim 125$ ns.\\

We experimentally verified our single- and two-qubit unitaries by Randomized Benchmarking (RB) \cite{Gambetta2012, Corcoles2013}. The following table shows the RB results for all single-qubit gates used in our experiments, including individual and simultaneous RB.

\begin{table}[h!]
	\centering
	\begin{tabular}{||c | c c | c c||} 
		\hline\hline
		Qubit 	 & $Q_0$ (83 ns) & $Q_1$ (83 ns)& $Q_0$ (125 ns) & $Q_1$ (125 ns) \\ [0.5ex]
		label & ($\times 10^{-3}$) & ($\times 10^{-3}$) & ($\times 10^{-3}$) & ($\times 10^{-3}$) \\
		\hline
		01 & - & $0.715\pm 0.005$& - & $1.244\pm 0.010$ \\ [0.5ex]
		\hline
		10 & $1.319\pm 0.017$ & - &$1.410\pm 0.010$ & - \\ [1ex]
		\hline
		11 & $1.367\pm 0.011$ & $0.763\pm 0.005$&$1.484\pm 0.014$ & $1.271 \pm 0.010$ \\ [0.5ex]
		\hline\hline
	\end{tabular}
	\caption{\label{table:SQRB} RB of our single-qubit gates. Qubit labels indicate which qubit was bencharmked on each case, with label $11$ indicating simultaneous RB.}
\end{table} 

Our two-qubit unitarias are CNOTs constructed from echo cross-resonance sequences \cite{Corcoles2013, Sheldon2016}. Each of the two cross-resonance pulses in a CNOT has durations of 333 and 500 ns for the two different gate lengths used in our experiments. For our two-qubit RB we obtain a CNOT error of $0.0373 \pm .0015$ ($0.0636 \pm .0021$) for the 333 (500) ns cross-resonance pulse.

\subsection*{Readout correction}
Our readout assigned fidelity was $\sim 95\%$ for both qubits.

For each experiment, we run 4 ($2^2$) calibration sequences preparing our two qubits in their joint computational states. We gather statistics of these calibrations and create a measurement matrix $A_{ij} = P(|i\rangle | |j\rangle)$ where $P(|n\rangle | |m\rangle)$ is the probability of measuring state $|m\rangle$ having prepared state $|n\rangle$. We then correct the observed outcomes of our experiments by inverting this matrix and multiplying our output probability distributions by this inverse.

\subsection*{Support vectors}

Here we show the support vectors and $\alpha_i$ as calculated for each of the three datasets studied from their $K$ matrices.

\begin{table}[!th]
	\begin{tabular}{|| c| c | c ||} 
		\hline\hline
		\multicolumn{3}{||c|}{Set I} \\
		\hline\hline
		SV  & $\alpha$& $y$   \\ [0.5ex]
		\hline
		[5.3407,2.9531]& 2.4185&+1 \\ [1ex]
		\hline
		[5.1522,4.6496]& 2.6248&+1 \\ [1ex]
		\hline
		[0.6912,0.7540]& 3.1379&+1 \\ [1ex]
		\hline
		[6.0947,5.7177]& 1.2891&+1 \\ [1ex]
		\hline
		[6.0947,4.1469]& 1.3604&+1 \\ [1ex]
		\hline
		[2.6389,4.8381]& 1.7163&+1 \\ [1ex]
		\hline
		[4.1469,2.0735]& 1.3242&-1 \\ [1ex]
		\hline
		[0.2513,3.2044]& 0.9757&-1 \\ [1ex]
		\hline
		[2.3248,3.9584]& 1.4953&-1 \\ [1ex]
		\hline
		[4.1469,6.2204]& 2.4536&-1 \\ [1ex]
		\hline
		[2.6389,1.6965]& 0.5293&-1 \\ [1ex]
		\hline
		[5.4664,4.6496]& 4.0213&-1 \\ [1ex]
		\hline
		[4.1469,3.0788]& 1.7477&-1 \\ [1ex]
		\hline
		\multicolumn{3}{||c|}{bias = -0.1090} \\
		\hline
		\hline
	\end{tabular}
	\quad
	\begin{tabular}{|| c| c | c ||} 
		\hline\hline
		\multicolumn{3}{||c|}{Set II} \\
		\hline\hline
		SV  & $\alpha$& $y$   \\ [0.5ex]
		\hline
		[0.1885,3.3301]& 2.1871&+1 \\ [1ex]
		\hline
		[5.4035,1.7593]& 1.443&+1 \\ [1ex]
		\hline
		[5.2150,3.7071]&2.0330&+1 \\ [1ex]
		\hline
		[4.71240,4.6496]& 0.7559&+1 \\ [1ex]
		\hline
		[0.4398,2.388]&1.0022&+1 \\ [1ex]
		\hline
		[6.2204,5.7177]& 2.3090&+1 \\ [1ex]
		\hline
		[5.9062,4.2726]& 1.6297&-1 \\ [1ex]
		\hline
		[1.8221,2.1362]& 2.9281&-1 \\ [1ex]
		\hline
		[0.3770,3.8327]& 3.0627&-1 \\ [1ex]
		\hline
		[0.31416,1.6965]& 1.3793&-1 \\ [1ex]
		\hline
		[1.6336,3.2672]& 0.2759&-1 \\ [1ex]
		\hline
		[2.5761,3.2673]& 0.4544&-1 \\ [1ex]
		\hline
		\multicolumn{3}{||c|}{bias = 0.2102} \\
		\hline
		\hline
	\end{tabular}
	\quad
	\begin{tabular}{|| c| c | c ||} 
		\hline\hline
		\multicolumn{3}{||c|}{Set III} \\
		\hline\hline
		SV  & $\alpha$& $y$   \\ [0.5ex]
		\hline
		[5.4664,3.7071]& 6.2649&+1 \\ [1ex]
		\hline
		[1.5080,5.2779]& 0.7777&+1 \\ [1ex]
		\hline
		[5.9690,3.8956]& 1.5619&+1 \\ [1ex]
		\hline
		[5.3407,3.5186]& 0.6482&+1 \\ [1ex]
		\hline
		[5.7177,4.5239]& 1.6387&+1 \\ [1ex]
		\hline
		[0.0628,3.5186]& 0.9379&+1 \\ [1ex]
		\hline
		[6.1575,1.445]& 0.3882&-1 \\ [1ex]
		\hline
		[5.9690,0.9425]& 0.8346&-1 \\ [1ex]
		\hline
		[0.2513,1.3823]& 2.5561&-1 \\ [1ex]
		\hline
		[5.2779,2.0735]& 4.8076&-1 \\ [1ex]
		\hline
		[6.0947,2.0106]& 3.2428&-1 \\ [1ex]
		\hline
		\multicolumn{3}{||c|}{bias = -0.1865} \\
		\hline
		\hline
	\end{tabular}
	\caption{\label{table:SV3s} Suport vectors for all three data sets used for our kernel estimation method, as shown as green circles in Figs. \ref{Figure3} (b) and \ref{FigureS1}.}
\end{table} 

\begin{figure}[h]
	\begin{center}
		\includegraphics[width=\textwidth]{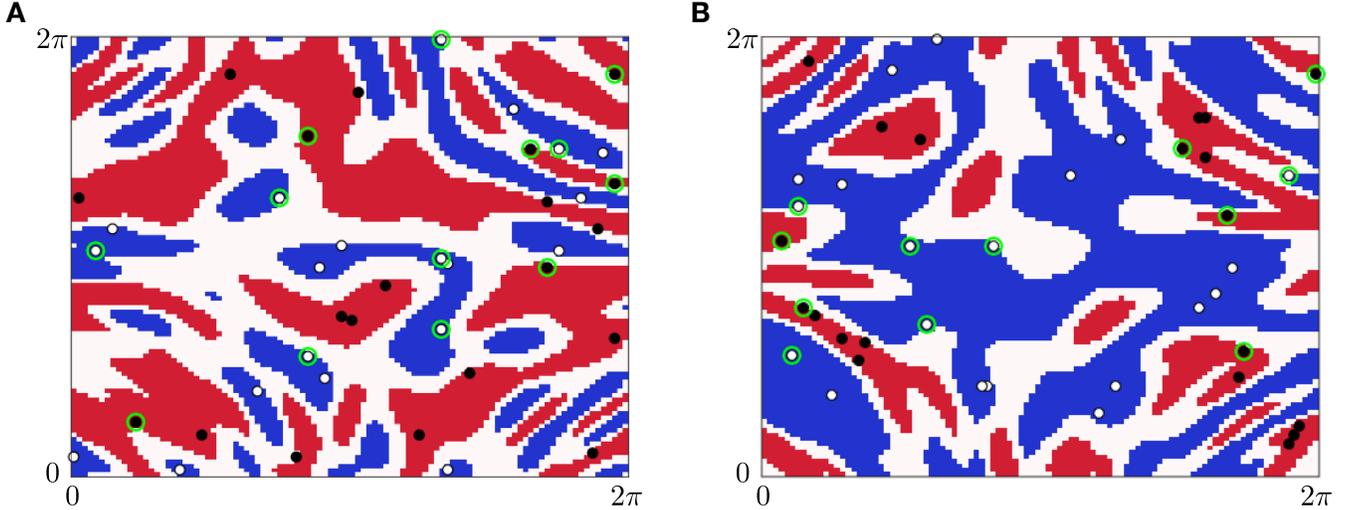}
		\caption{ \label{FigureS1} Sets I (a) and II (b), including training data points (white and black circles) and support vectors (green circles).}
	\end{center}
\end{figure}

\begin{figure}[h]
	\begin{center}
		\includegraphics[width=\textwidth]{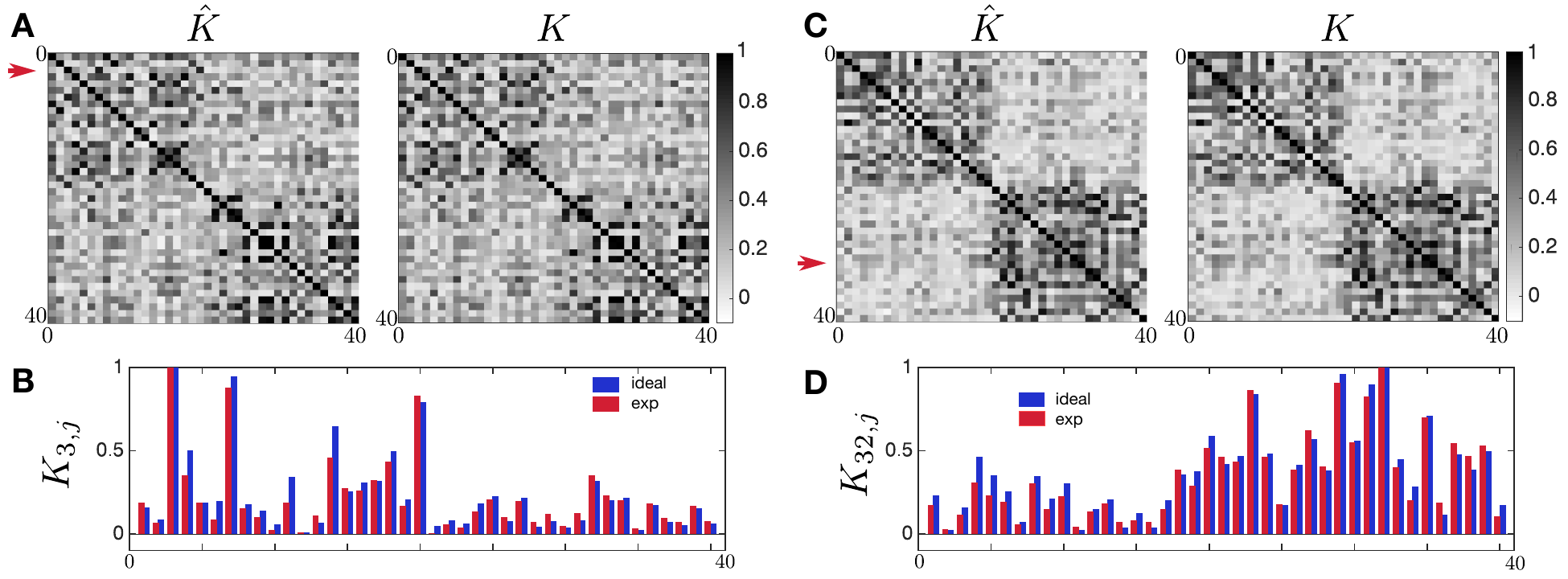}
		\caption{ \label{FigureS2} Experimentally estimated and ideal kernel matrices for Sets I (a, b) and II (c, d). For both datasets we show a cut (b, d) through the row at which the maximum of $|\hat{K}-K|$ occurs, similarly as in Fig. \ref{Figure4} in the main text.}
	\end{center}
\end{figure}

\subsection*{Error mitigation for kernel estimation}
The experimental estimation of the kernel matrices shown in Fig. \ref{FigureS2} and in Fig. \ref{Figure4} in the main text involves running the experiments at different gate lengths and extrapolating the expectation value of the observable of interest to its zero-noise value. While this technique can be extremely powerful in scenarios where the noise is invariant under time rescaling, it is particularly sensitive to measurement sampling noise. In many cases it is the experimental readout assignment fidelity that determines the bound on how precisely the observable can be estimated. 

Even though for our Sets I and II we attain 100 \% classification success over 10 randomly drawn test sets in each case, we can quantify how close our experimentally determined separating hyperplane is to the ideal. 

The optimal hyperplane for a given training set can be expressed as the linear combination $\sum_{i} \alpha_i y_i \vec{x}_i  = {\bf w}$ (eqn. \ref{w_dl}), where $\bf{w}$ is a vector orthogonal to the optimal separating hyperplane and $\vec{x}_i$ are the support vectors. We can therefore quantify the distance between the experimentally obtained hyperplane and the ideal hyperplane by calculating the inner product $\langle {\bf{w}}, {\bf{w}_{\textrm{ideal}}} \rangle = \sum_{i \in N_S}\sum_{j \in N_S'} y_i y_j \alpha_i^* \alpha_j |\langle \vec{x}_i \vec{x}_j\rangle|^2/||\bf{w}|| ||\bf{w}_{\textrm{ideal}}||$ where $N_S$ and $N_S'$ are the sets of experimentally obtained and ideal support vectors, respectively.

\begin{figure}[h]
	\begin{center}
		\includegraphics[width=0.5\textwidth]{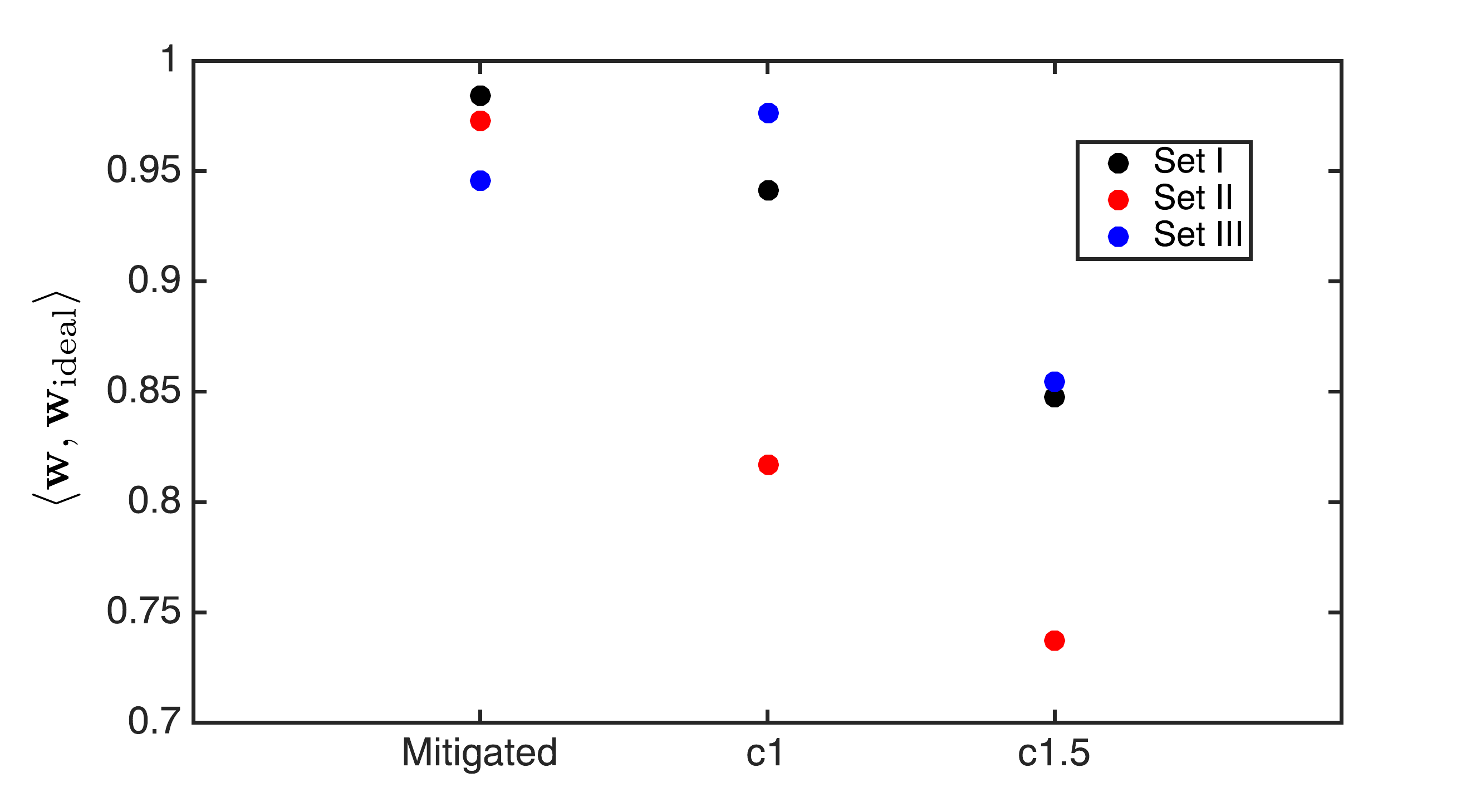}
		\caption{ \label{FigureS8} Inner products between experimentally obtained hyperplanes and the ideal for each training set. The $x-$axis shows the results for the two different time stretches used in our experiments, $c1$ and $c1.5$, corresponding to the faster and slower gates, respectively. We also show the error-mitigated hyperplanes inner products.}
	\end{center}
\end{figure}

In Fig.~ \ref{FigureS8} we show the inner products between the ideal and all experimental hyperplanes, including the two sets of gate times used throughout our experiments, $c1$ and $c1.5$, as well as the error-mitigated hyperplanes.

For Sets I and II, which classify at 100 \% success, it is clear that error mitigation improves our results very significantly. This is not the case for Set III, which classifies at 94.75 \% success. In fact, for Set III we see that error-mitigation worsens the hyperplane, as the results are closer to the ideal for the unmitigated experiments than both Sets I and II. 

A look at the calibration data taken along the direct kernel estimation experiments for each set, we see that the readout assignment fidelities of $Q_0$($Q_1$) are 96.56\% (96.31\%) for Set I, 95.90\% (96.36\%) for Set II, and 93.99\% (95.47\%) for Set III. The slightly worse readout fidelities for Set III could partially explain the worse classification results for this set, but other aspects of the protocol might also contribute to this, like for example some gates operating on a somewhat non-linear regime after the calibrations in this set.

Another symptom for the degree of classification success in a given set can be observed by looking at the combined weight of negative eigenvalues in the kernel matrix,  $\sum_{k_i<0} |k_i|$, with $k_i$ the eigenvalues of the kernel. We obtain 1.40, 1.27 and 2.41 for Sets I, II, and III, respectively.
\end{document}